\documentclass[journal]{IEEEtran}

\usepackage{cite}

\ifCLASSINFOpdf
  \usepackage[pdftex]{graphicx}
  \graphicspath{{figures/}}
\else
\fi

\usepackage[cmex10]{amsmath}
\usepackage{array}

\usepackage{url}

\usepackage{algorithm}  %
\usepackage{algpseudocode}
\usepackage{multirow}
\usepackage{amssymb}
\usepackage{mathtools}
\usepackage[hidelinks]{hyperref}
\usepackage{tikz}
\usepackage{microtype}

\newcounter{solver}
\newenvironment{solver}{\refstepcounter{solver}\equation}{\tag{$S_\thesolver$}\endequation}

\definecolor{annotred}{rgb}{0.59,0.157,0.078}
\definecolor{labelgray}{rgb}{0.2,0.2,0.2}

\def\R{{\mathbb R}}

\def\dict{{P}}
\def\act{{A}}
\def\actB{{B}}
\def\actG{{G}}
\def\spec{{V}}
\def\copyact{{\overline{A}}}
\newcommand{\apspecvar}[1]{{\dict{#1}}}
\def\apspec{{\apspecvar{\act}}}
\def\apspecBG{{\dict(\actB \odot\actG)}}
\def\dist{{D}}
\def\distsmall{{d}}
\def\objf{{f}}
\def\tobjf{{h}}
\def\charfunc{{\chi}}

\newcommand{\ind}[1]{{(#1)}}
\DeclareMathOperator*{\argmin}{argmin}
\DeclareMathOperator*{\argmax}{argmax}
\DeclareMathOperator{\sign}{sign}
\DeclarePairedDelimiter{\normcore}{\lVert}{\rVert}
\newcommand{\norm}[1]{\normcore[\big]{#1}}
\def\markov{{\mathcal M}}
\def\threshset{{\mathcal T}}

\def\markov{{\mathcal M}}
\def\minact{{a_m}}
\def\tdvop{{\Delta_D}}
\def\xlinadmm{{X^{k+1}}}  %

\begin{document}
\title{Piano Transcription in the Studio Using an Extensible Alternating Directions Framework}

\author{Sebastian~Ewert,~\IEEEmembership{Member,~IEEE,}
        and~Mark~Sandler,~\IEEEmembership{Fellow,~IEEE}%
\thanks{S.~Ewert and M.~Sandler are with Queen Mary University of London, UK.}%
\thanks{Manuscript received Nov 2015; revised Apr 2016.}} %

\markboth{IEEE/ACM Transactions on Audio, Speech, and Language Processing,~Vol.~?, No.~?, Month~2016}%
{Ewert \MakeLowercase{\textit{et al.}}: Piano Transcription in the Studio Using an Extensible Alternating Directions Framework}

\maketitle

\begin{abstract}
Given a musical audio recording, the goal of automatic music transcription is to determine a score-like representation of the piece underlying the recording. Despite significant interest within the research community, several studies have reported on a ``glass ceiling'' effect, an apparent limit on the transcription accuracy that current methods seem incapable of overcoming.
In this paper, we explore how much this effect can be mitigated by focusing on a specific instrument class and making use of additional information on the recording conditions available in studio or home recording scenarios. In particular, exploiting the availability of single note recordings for the instrument in use we develop a novel signal model employing variable-length spectro-temporal patterns as its central building blocks -- tailored for pitched percussive instruments such as the piano.
Temporal dependencies between spectral templates are modeled, resembling characteristics of factorial scaled hidden Markov models (FS-HMM) and other methods combining Non-Negative Matrix Factorization with Markov processes. In contrast to FS-HMMs, our parameter estimation is developed in a global, relaxed form within the extensible alternating direction method of multipliers (ADMM) framework, which enables the systematic combination of basic regularizers propagating sparsity and local stationarity in note activity with more complex regularizers imposing temporal semantics. The proposed method achieves an f-measure of 93-95\% for note onsets on pieces recorded on a Yamaha Disklavier (MAPS DB). 
\end{abstract}


\IEEEpeerreviewmaketitle

\section{Introduction}
\label{sec:Intro}

\IEEEPARstart{A}{utomatic} Music Transcription (AMT) has a long history in music processing \cite{Moorer77_Transcription_CMJ}. 
Identifying higher-level musical concepts such as notes in digital music recordings, it is often considered a key technology for a semantic analysis of music,
with applications ranging from various retrieval tasks in music informatics over computational musicology and performance analysis to creative music technology \cite{KlapuriD06_SPforMusic_BOOK}.
Despite significant interest within the research community, transcription remains highly challenging and accuracies reported in recent years seem to have reached a plateau %
\cite{BenetosDGKK13_AMT-Future_JIIS}.
A promising approach to achieve higher accuracy is to provide a method with additional prior information. %
For example, the user can be actively involved in the transcription process providing partial transcription results to stabilize the parameter estimation process \cite{KirchhoffDK12_ShiftVariantSemiAutomNMF_ISMIR}.
While this is an interesting direction and leads to measurable improvements in accuracy, it can be time-consuming and precludes a fully automatic process.

A central goal of this paper is to investigate if the performance of a transcription system can be improved by focusing on a specific application scenario.
First, we focus on a single class of instruments, namely pitched percussive instruments such as the piano or harpsichord.
Second, we assume that recordings of single notes are available for the instrument to be transcribed (at least one playing style),
which mitigates many uncertainties regarding the instrument and the recording conditions (room response, recording equipment). %
Further, to obtain a practical transcription system, we assume that the user can play at the beginning of a recording session a note in pianissimo (low intensity), which is used by our system to derive a threshold employed to differentiate between an active note and estimation noise.
Given this scenario, we can tailor our proposed signal model to precisely this instrument class, which is necessary to account for the highly non-stationary behavior of the piano sound production process.
In particular, when a key on the piano is hit a mechanical, pitch-dependent, broad-band sound is produced, followed by a harmonic sound that is non-stationary due to amplitude modulation (\emph{beating}) and differences in decay rate between the harmonics \cite{FletcherR91_instruments}.
The main idea in our signal model is that this sound sequence is highly characteristic for a note event, which enables us to use the single note recordings as a blueprint to identify similar spectro-temporal patterns in the recording. 

To implement this idea, existing methods are not well suited. In particular, most state of the art transcription methods employ variants of \emph{non-negative matrix factorization (NMF)}, which treat spectral and temporal information independently and therefore cannot make use of such joint patterns (see also Section~\ref{sec:Related}). As an extension to NMF, \emph{Non-Negative Matrix Deconvolution (NMD)} employs spectro-temporal patterns -- however, these have a fixed length and thus are not a good match for modeling notes of variable duration. 
A promising candidate %
could be the \emph{factorial scaled hidden Markov model (FS-HMM)} \cite{OzerovFC09_FSHMM_WASPAA} and its variants \cite{MysoreS12_VariationalNFHMM_ICML,MasahiroLKNOS11_NonParametricFactorialHMM_WASPAA}, in which a Markov process governs which spectral templates can be used in a given time frame based on the previous frame.
Such models, however, were proposed in the context of modeling a few concurrent speakers, and, in a standard form, their computational complexity is exponential in the number of Markov processes\cite{MysoreS12_VariationalNFHMM_ICML}, of which we have $88$ in our case (one for each piano key).
Parameter decoupling techniques such as generalized expectation maximization often used in this context (i.e. fixing some parameters while optimizing against the remaining ones) tend to converge extremely slowly with so many independent processes, and often lead to poor local minima with excessive decoupling (see also \cite{GhahramaniJ97_FactorialHMM_ML,MysoreS12_VariationalNFHMM_ICML,EwertPS15_DPNMD_ICASSP}).

The method presented in this paper employs spectro-temporal patterns of variable length to model a given time-frequency representation of a piano recording.
The core idea for our parameter estimation stems from the observation that piano sounds can be well represented using simple left-to-right Markov models (i.e. there is a clear succession of spectral templates when a piano key is hit). As we will see, this enables us, instead of strictly enforcing Markov properties as in FS-HMMs,
to approximate the temporal transitions between spectral templates in a relaxed form by stating the parameter estimation problem as a structured sparse coding problem, which is controlled by simple convex regularizers.
Using these regularizers we can steer the solution close to a semantically meaningful progression similar to an FS-HMM solution.
The resulting problem is convex and all parameters are jointly optimized (i.e. no decoupling is necessary), such that poor local minima are typically avoided.
Once we are close to a solution to this convex problem, we switch to non-convex regularizers, which can be interpreted as projections onto matrices encoding strict Markov-like transitions. 
While these non-convex regularizers would lead to relatively poor results without the convex initialization, we can use this combination to further refine the parameter estimate, %
gradually enforcing stricter, more meaningful transitions between spectral templates.
The entire model is developed within the highly extensible \emph{alternating direction method of multipliers (ADMM)} framework, which is widely used in the sparse-coding and computer vision communities but has not yet received the same amount of attention in audio processing research despite its proven usefulness in non-linear optimization of non-differentiable functions and machine learning for big data.
	
The remainder is organized as follows. In Section~\ref{sec:Related} we discuss related work, focusing on core concepts in current transcription methods. In Section~\ref{sec:Model}, we describe the proposed model and explain the effects of specific regularizers. Next, in Section~\ref{sec:parameterEst}, we develop an efficient parameter estimation method for our model based on the ADMM algorithmic framework, which we describe in more detail as we think it might be useful in other contexts as well. In Section~\ref{sec:Experiments}, we discuss the results of various experiments to illustrate the performance of the proposed model as well as the influence of parameters. Finally, in Section~\ref{sec:Conclusions} we conclude the paper with an outlook on future work.

\section{Related Work}
\label{sec:Related}

As one of the central topics in music processing, automatic music transcription (AMT) has attracted considerable interest within the research community over the years.
In the following, we refer to overview articles for a more comprehensive overview and focus on discussing central contributions and general concepts.
In particular, many methods proposed before $2005$ are described in \cite{KlapuriD06_SPforMusic_BOOK}, and many more recent methods in an outlook article \cite{BenetosDGKK13_AMT-Future_JIIS}.

Overall, a wide range of strategies have been used for AMT. For example, in \cite{BarbanchoBJ04_PianoTranscription_AA,Cemgil04_BayesianAMT_PHD} the recording to be transcribed is first segmented by detecting onsets and rhythmic structures, which is then used to guide a subsequent pitch estimation. A large body of work involves elaborate methods exploiting the harmonicity of musical sounds to group detected spectral peaks into note objects \cite{Dixon00_PianoTrans_AMCM,BelloDS06_TimeDomainInformation_TASLP,DuanPZ10_MultiF0_TASLP,YehRR10_MultiPitch_TASLP}.
Another group of methods employs probabilistic sinusoidal plus noise modeling, in which parameters such as onset and offset position, fundamental frequency, intensity and spectral envelope parameters are adjusted to match the observed signal using maximum a posteriori estimation \cite{EmiyaBD10_MultipitchEstimation_TASLP} or genetic algorithms \cite{ReisDF12_AMTGenetic_TASLP}.
Another approach employs a hidden Markov model (HMM) \cite{Raphael02_TranscriptionPiano_ISMIR}, where each state corresponds to one possible combination of active notes, which requires elaborate heuristic state-space pruning strategies to be computationally feasible. 

A further successful technique is based on the iterative estimation and subtraction of the predominant fundamental frequency and its corresponding harmonics \cite{Klapuri08_pitch_IEEETASL}.
Transcription has also been considered as a classification or regression task in the context of discriminative methods. In \cite{PolinerE07_PolyphonicPiano_EURASIP}, a total of $87$ support vector machines were trained to detect pitch activity in spectrogram frames, followed by temporal smoothing using an HMM. As another example, the system presented in \cite{Marolt04_AMT-NN-Piano_TMM} was highly successful in comparative studies \cite{BertinBV10_EnforcingHarmonicityInBayesNMF_TASLP} and uses time-delay neural networks (similar to convolutional networks in time direction) to classify the output of adaptive oscillators, which are used to track and group partials in the output of a gammatone filterbank. More recently, \cite{BoeckS12_RNN-Transcription_ICASSP} presented a system using a multi-layer recurrent neural network based on Hochreiter and Schmidhuber's bidirectional long short term memory units, which were chosen to better model temporal dependencies in music. 

Most state of the art methods, however, employ variants of non-negative matrix factorization (NMF), see \cite{VirtanenGRS15_CompositionalModels_IEEESPM} for a recent overview.
In general, the underlying idea of NMF-based methods is to model a given time-frequency representation of a recording as a mixture of note- or sound-specific spectral template vectors and to estimate their individual
activity over time. One principal advantage of NMF over many early approaches is that parameters in NMF are iteratively refined and errors made early in the process can thus be corrected later.
Further, many NMF-based methods contain parameters with a clear interpretation, which can facilitate the integration of prior knowledge and thus can give an advantage over some discriminative models where this can be more difficult \cite{OzerovVB12_PriorInfoSourceSep_TASLP}. 
For example, many recent approaches have extended classic NMF \cite{SmaragdisB03_MusicTranscriptionNMF_WASPAA} by forcing the spectral templates to represent only harmonic sounds.
In particular, the methods presented in \cite{BertinBV10_EnforcingHarmonicityInBayesNMF_TASLP,FuentesBR11_ShiftPLCAForMultiF0_ICASSP} restrict the spectral templates to a linear combination of narrow-band harmonic sub-templates, each having a clear pitch association. As another example, the PreFest \cite{Goto04_Prefest_ISCA} and HTC \cite{KameokaNS07_MultipitchAnalyzer_TASLP} methods can be interpreted as modeling spectral templates using a set of scaled Gaussians, each representing a partial of a harmonic sound in frequency direction. %
To transcribe recordings of several instruments, a typical approach is to use pre-trained spectral templates for specific instruments \cite{GrindlayE09_AMTEigenInstr_WASPAA,BenetosEW14_PitchUnpitchedTranscription_ICASSP}. %
Further, a high number of spectral templates can be used to increase the representation accuracy of the model, which leads to sparse coding methods \cite{AbdallahP06_TranscriptionSparseCoding_TNN} where the model is encouraged to explain the audio using only a few of the available templates, see also \cite{OhanlonNP12_StructuredSparsityAMT_ICASSP} for a recent study in an AMT context.

An overarching principle of NMF-based methods is that spectral properties are decoupled from temporal ones, i.e. neither do the activations provide information about how a note spectrally manifests nor do the templates describe when a note occurs or how it evolves.
Without an enforced temporal progression of spectral templates, however, non-stationary signals like a piano note are difficult to model. %
For example, one typically cannot express in NMF that a certain spectral template for the sustain part of a note is expected after a certain time after the attack.
Further, activations between neighboring frames are often not or only loosely coupled, such that it is difficult to express that an activation value has a certain relationship to the ones in subsequent frames.

An extension to NMF modeling such spectro-temporal dependencies was presented in \cite{Smaragdis04_NMD} under the name \emph{Non-Negative Matrix Deconvolution (NMD)}, which uses, instead of spectral templates, entire time-frequency patterns concatenating several templates over time as building blocks within the model. The use of patterns enforces a specific temporal order for the templates and effectively couples their activations. 
However, since these patterns have a fixed length, NMD has not been used to transcribe instruments such as the piano, where notes are of variable duration. 
The FS-HMM approach \cite{OzerovFC09_FSHMM_WASPAA,MysoreS12_VariationalNFHMM_ICML,MasahiroLKNOS11_NonParametricFactorialHMM_WASPAA} adds more flexibility regarding the length of template sequences by employing a Markov process that governs the use of specific templates in a given frame. While such models indeed provide the necessary freedom to model a non-stationary sound such as a piano note, their computational complexity in a basic formulation is typically exponential in the number of Markov processes \cite{MysoreS12_VariationalNFHMM_ICML} -- with $88$ piano keys and corresponding Markov processes this is computationally infeasible.
Updating the parameters of some Markov processes while keeping the remaining ones fixed (i.e. applying generalized expectation maximization schemes) typically converges extremely slowly with so many processes and often leads to poor local minima in the distance function between model and observed signal \cite{GhahramaniJ97_FactorialHMM_ML} (see \cite{BenetosD13_MusicTranscription_JASA} in a transcription scenario).
A hybrid between NMD and FS-HMM for piano transcription was presented in \cite{EwertPS15_DPNMD_ICASSP}, where the computational issues are approached by a combination of decoupling strategies similar to Viterbi training to generate note event candidates and global, coupled optimization over all candidates during an activation update.
While this type of parameter estimation typically yields transcription results of comparatively high quality, the remaining decoupling during the candidate selection can still lead to some poor local minima, which limits the transcription performance, as we will see below.

\section{Proposed Model}
\label{sec:Model}
While the goal of our proposed method is similar to the approach presented in \cite{EwertPS15_DPNMD_ICASSP}, the underlying model and parameter estimation process are fundamentally different.
The design goals were to eliminate the decoupling of parameters, and as we will see, the resulting model is not only simpler but also yields improved results and is computationally more efficient.
In the following, we assume that for each of the $K=88$ piano keys a recording of a single note for the instrument to be transcribed is available.
Computing a log-frequency magnitude spectrogram from each recording, we obtain $K$ time-frequency patterns, each consisting of $L$ spectral templates, resulting in a
\emph{pattern dictionary tensor} $\dict \in \R_{\ge 0}^{M \times L \times K}$, where $M$ is the number of frequency bins. 
Each column $\dict\ind{:,\ell,k} \in \R_{\ge 0}^{M}$ for fixed $\ell$ and $k$ contains a single \emph{spectral template vector} or \emph{template} for short; here we used the Matlab notation \emph{:} to refer to all elements in an index dimension, i.e. $\{1,\ldots,M\}$ in this case.
In contrast to most NMF approaches, we do not normalize the templates in order to preserve their energy progression over time, which later will provide an additional indication for which part of the note pattern should be active in a given time frame.

Next, given a log-frequency magnitude spectrogram
$\spec \in \R_{\ge 0}^{M \times N}$ of a recording to be transcribed, we model each entry in $\spec$ as a sum over the $K$ patterns. More precisely:
\begin{eqnarray}
\spec\ind{m,n} \approx (\apspec)\ind{m,n} := \sum_k \sum_\ell \dict\ind{m,\ell,k} \cdot \act\ind{k,\ell,n},
\label{eq:mainModel1}
\end{eqnarray}
where $\act \in \R_{\ge 0}^{K \times L \times N}$ is the \emph{activity tensor}, which specifies the intensity of each template in each frame.
Our goal will be to design and minimize a function of $\act$ describing a semantically meaningful distance (or divergence) between $\spec$ and $\apspec$.
The final distance function will consist of several terms each encouraging certain behavior in $\act$ in a soft way by penalizing unwanted behavior. 
That means that our model does not directly impose any structure on $\act$, which is in stark contrast to many NMF approaches discussed in Section~\ref{sec:Related}.
For example, a temporal order of templates is not hardly enforced within the model as in more complex approaches such as FS-HMM or in \cite{EwertPS15_DPNMD_ICASSP}.
This lack of enforcement will enable us to build efficient and not hardly decoupling parameter estimation procedures in Section~\ref{sec:parameterEst}.

\begin{figure*}
		\hspace{-2mm}
    \begin{tikzpicture}[scale=1.0, every node/.style={transform shape}]
			\begin{scope}[shift={(0,1.5)}]
			\node[anchor=south west] (label) at (0.3,6){\includegraphics[width=3.7cm,height=1cm]{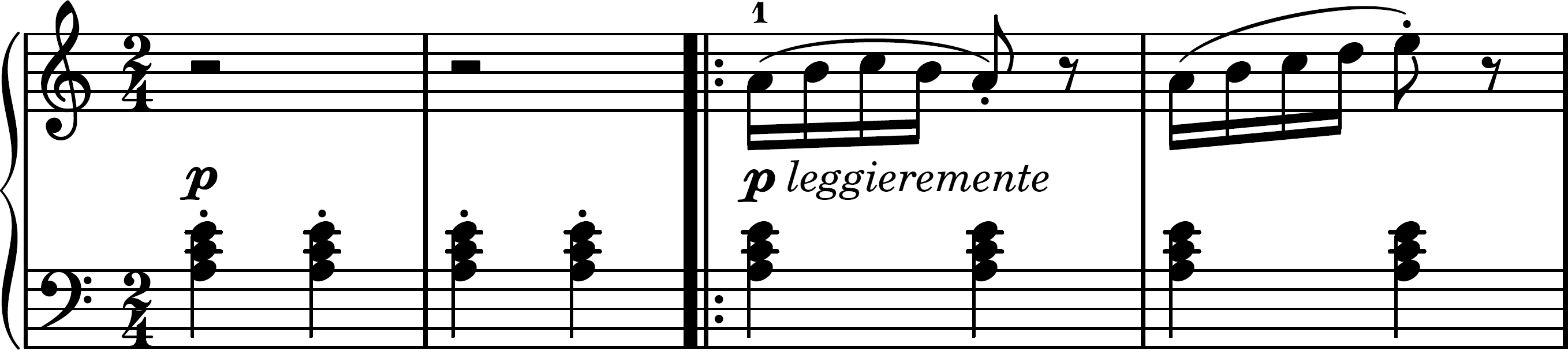}};
			\node[anchor=south west] (label) at (0.35,0){\includegraphics[width=3.65cm]{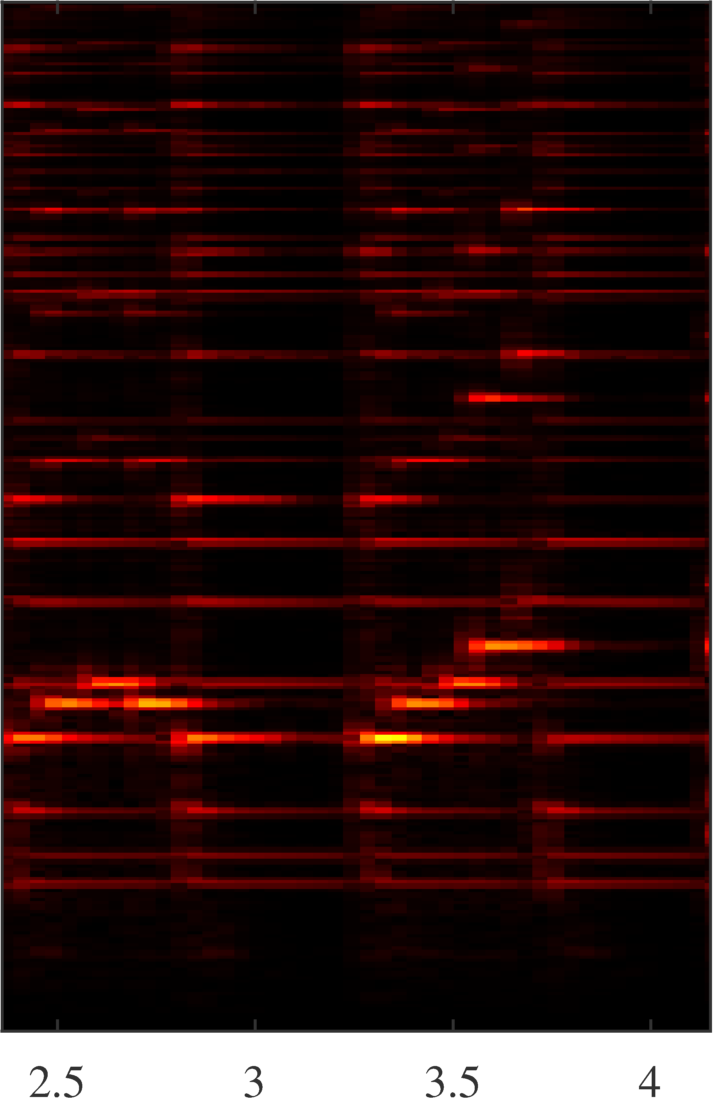}};
			\node at (2.3,-0.1) {\footnotesize Time [sec]};
			\node (label) at (2.3,7.4) {\small (a)}; 
			\end{scope}

			\begin{scope}[shift={(4.6,5.1)}]
      \node[anchor=south west] (label) at (0,0){\includegraphics[width=4cm,height=4.3cm]{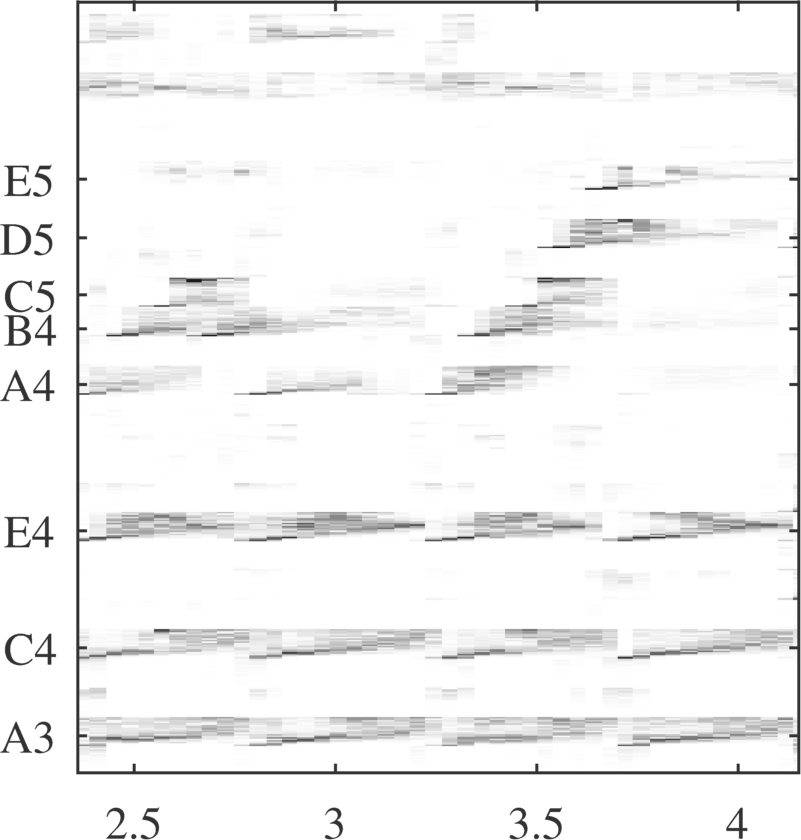}};
			\node[anchor=south west,rotate=90] at (0.2,1.2) {\scriptsize \textcolor{labelgray}{Spectral Template ID}};
			\node at (2.3,-0.1) {\scriptsize \textcolor{labelgray}{Time [sec]}};
			\node (label) at (2.3,4.65) {\small (b)}; 
			\draw[annotred,thick,rounded corners=2pt] (2.2,2.25) rectangle (3.4,3.1);
			\draw[annotred,thick,rounded corners=2pt] (2.1,3.9) rectangle (4.15,4.15);
			\end{scope}
			\begin{scope}[shift={(9.2,5.1)}]
      \node[anchor=south west] (label) at (0,0){\includegraphics[width=4cm,height=4.3cm]{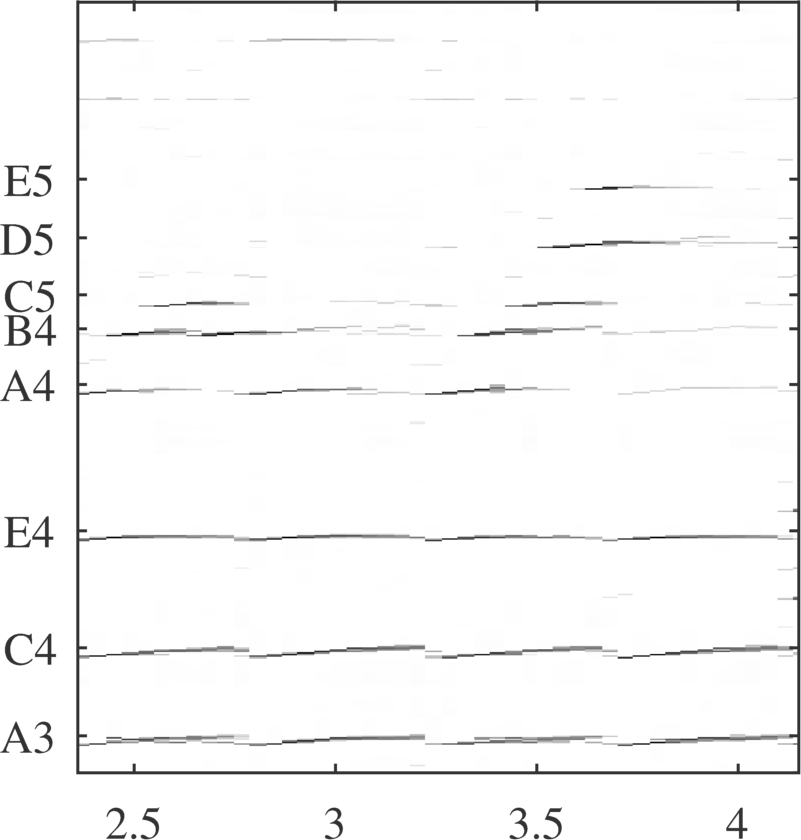}};
			\node[anchor=south west,rotate=90] at (0.2,1.2) {\scriptsize \textcolor{labelgray}{Spectral Template ID}};
			\node at (2.3,-0.1) {\scriptsize \textcolor{labelgray}{Time [sec]}};
			\node (label) at (2.3,4.65) {\small (c)}; 
			\draw[annotred,thick,rounded corners=2pt] (1.3,4.13) rectangle (2.2,4.3);
			\draw[annotred,thick,rounded corners=2pt] (0.48,1.50) rectangle (1.35,1.8);
			\end{scope}
			\begin{scope}[shift={(13.8,5.1)}]
      \node[anchor=south west] (label) at (0,0){\includegraphics[width=4cm,height=4.3cm]{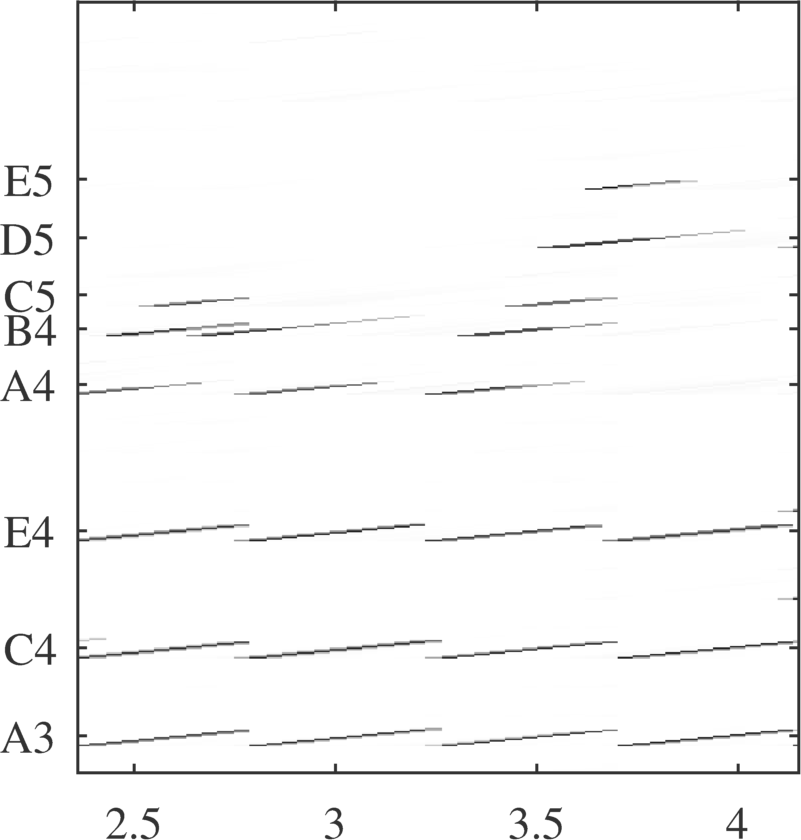}};
			\node[anchor=south west,rotate=90] at (0.2,1.2) {\scriptsize \textcolor{labelgray}{Spectral Template ID}};
			\node at (2.3,-0.1) {\scriptsize \textcolor{labelgray}{Time [sec]}};
			\node (label) at (2.3,4.65) {\small (d)}; 
		  \draw[annotred,thick,rounded corners=2pt] (0.9,2.63) rectangle (1.55,2.83);
			\draw[annotred,thick,rounded corners=2pt] (0.48,0.95) rectangle (0.8,1.25);
			\end{scope}
			
			\begin{scope}[shift={(4.6,0)}]
      \node[anchor=south west] (label) at (0,0){\includegraphics[width=4cm,height=4.3cm]{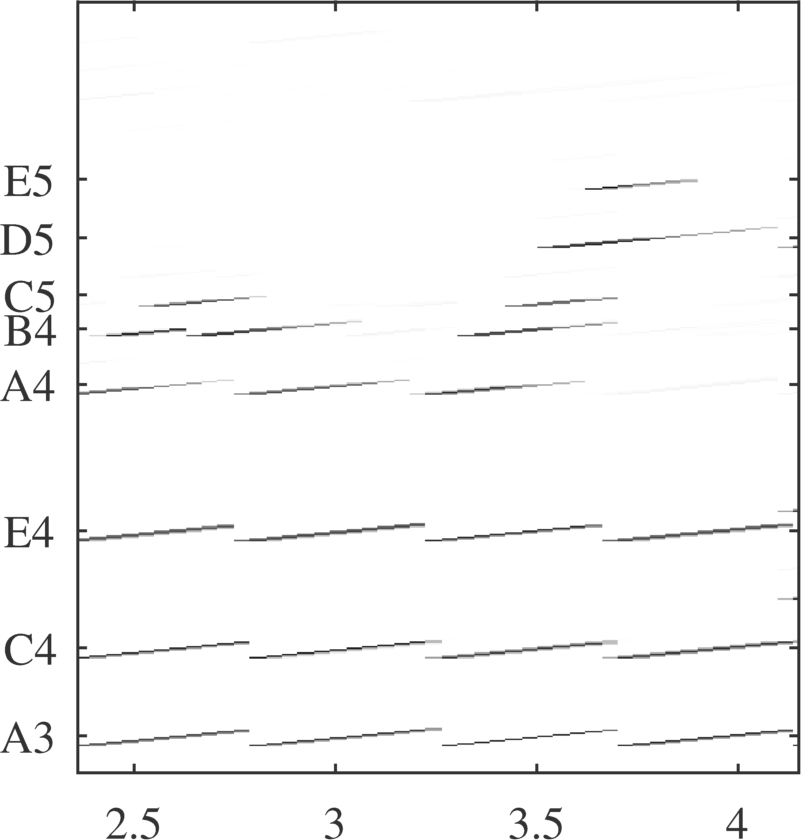}};
			\node[anchor=south west,rotate=90] at (0.2,1.2) {\scriptsize \textcolor{labelgray}{Spectral Template ID}};
			\node at (2.3,-0.1) {\scriptsize \textcolor{labelgray}{Time [sec]}};
			\node (label) at (2.3,4.65) {\small (e)}; 
			\draw[annotred,thick,rounded corners=2pt] (2.1,3.8) rectangle (4.0,4.15);
			\draw[annotred,thick,rounded corners=2pt] (3.1,2.25) rectangle (4.1,2.6);
			\end{scope}
			\begin{scope}[shift={(9.2,0)}]
      \node[anchor=south west] (label) at (0,0){\includegraphics[width=4cm,height=4.3cm]{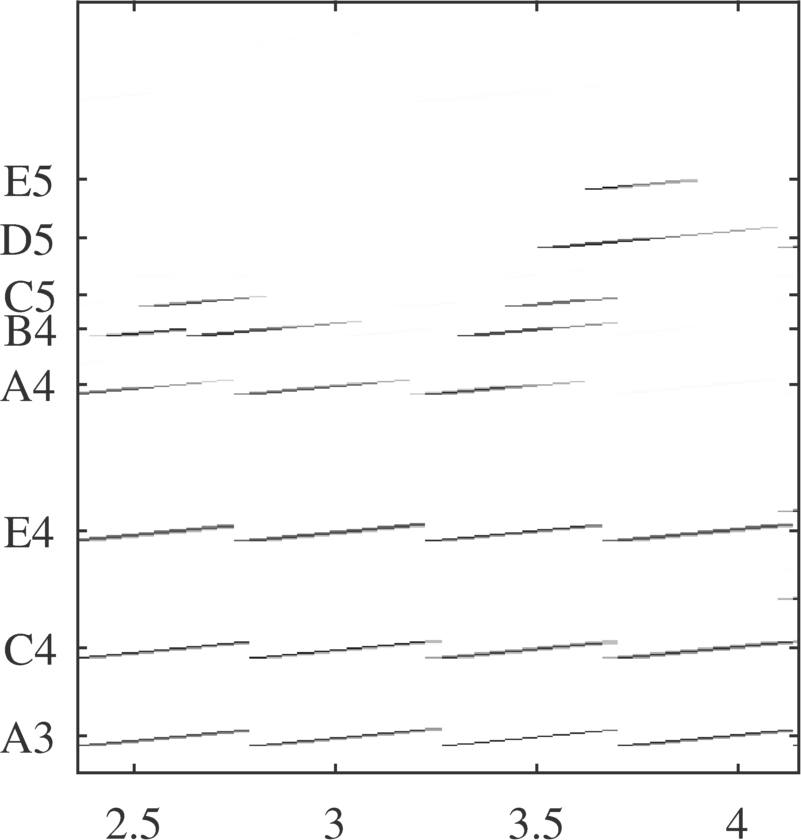}};
			\node[anchor=south west,rotate=90] at (0.2,1.2) {\scriptsize \textcolor{labelgray}{Spectral Template ID}};
			\node at (2.3,-0.1) {\scriptsize \textcolor{labelgray}{Time [sec]}};
			\node (label) at (2.3,4.65) {\small (f)}; 
			\draw[annotred,thick,rounded corners=2pt] (3.35,3) rectangle (4.1,3.4);
			\end{scope}
			\begin{scope}[shift={(13.8,0)}]
      \node[anchor=south west] (label) at (0,0){\includegraphics[width=4cm,height=4.3cm]{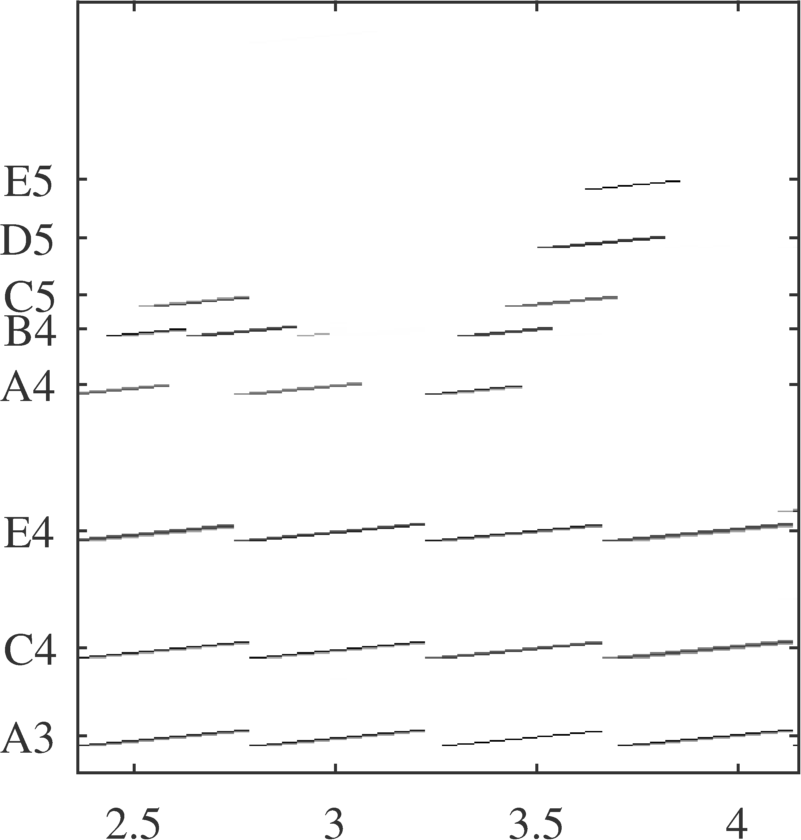}};
			\node[anchor=south west,rotate=90] at (0.2,1.2) {\scriptsize \textcolor{labelgray}{Spectral Template ID}};
			\node at (2.3,-0.1) {\scriptsize \textcolor{labelgray}{Time [sec]}};
			\node (label) at (2.3,4.65) {\small (g)}; 
			\draw[annotred,thick,rounded corners=2pt] (1.6,2.6) rectangle (1.8,2.8);
			\end{scope}
    \end{tikzpicture}
    \caption{\textbf{(a)} Log-frequency spectrogram of a recording of bars 3 and 4 of Burgm{\"u}ller's Opus 100 Etude 2. (b)-(g) Activity tensor estimated based on various objective functions and regularizers. \textbf{Convex terms}: \textbf{(b)} Kullback-Leibler and non-negativity term, \textbf{(c)} $\ell_1$ or LASSO regularizer, \textbf{(d)} total diagonal variation regularizer. \textbf{Non-convex terms}: \textbf{(e)} Markov-state regularizer, \textbf{(f)} threshold regularizer, \textbf{(g)} Binary Markov-state and strict coupling regularizer. Marked areas are discussed in the text.}
		\label{fig:IlluRegul}
\end{figure*}
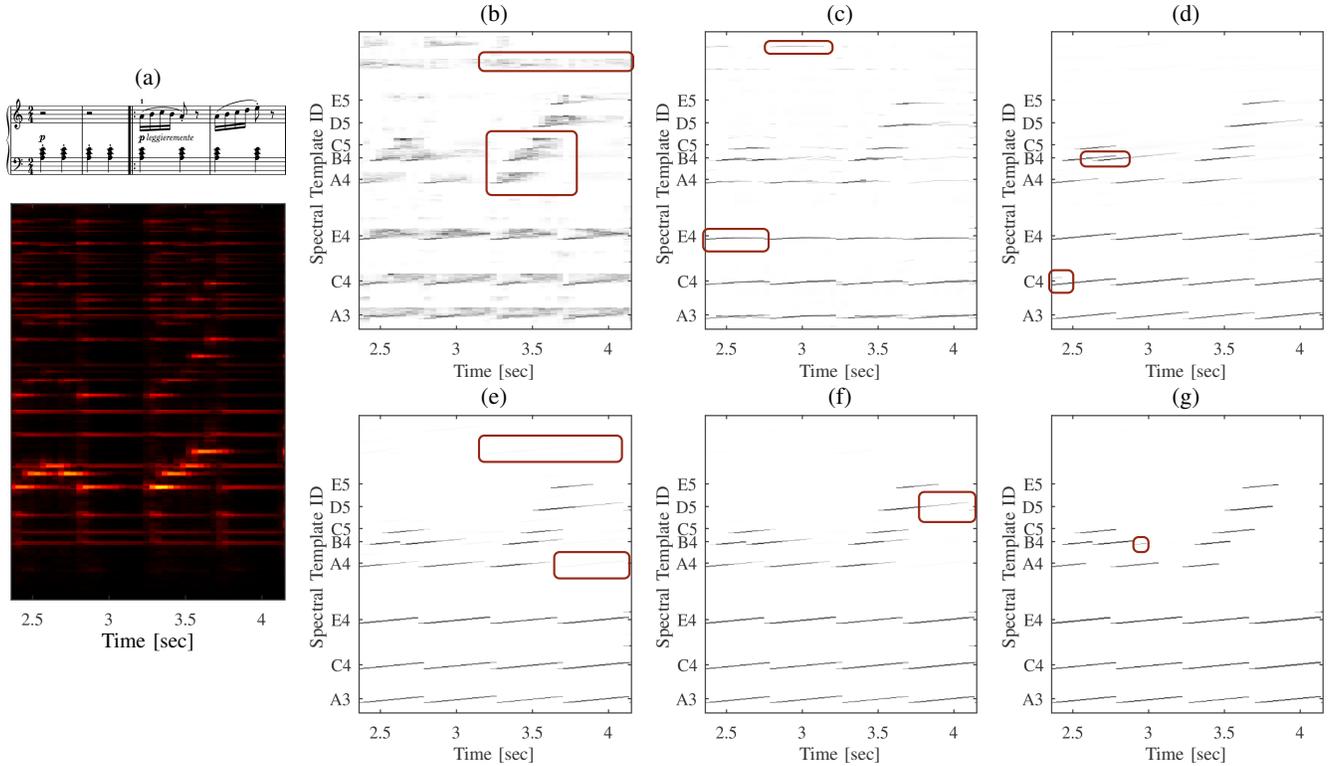

\subsection{Encouraging Data Fidelity and Non-Negativity}

As a first step, we include a data fidelity term in our distance function in the form of a generalized divergence between $\spec$ and $\apspec$. Here, most other sparse coding methods employ the Frobenius norm \cite{AbdallahP06_TranscriptionSparseCoding_TNN}. For NMF-based AMT methods, however, the use of other divergences led to improved results. We use the generalized Kullback-Leibler (KL) divergence
\[ \objf_1(\apspec) := \dist(\spec, \apspec) :=\sum_{m,n}\distsmall(\spec\ind{m,n},(\apspec)\ind{m,n})\]
with $\distsmall(a,b) := a \cdot \log\left(\frac{a}{b}\right) - a + b$ for $a,b> 0$. These improvements were often attributed to the observation that the differences between $\spec$ and $\apspec$ are distributed in practice rather according to a Poisson than a Gaussian distribution, which suggests the use of the KL divergence \cite{AbdallahP06_TranscriptionSparseCoding_TNN}. 
However, the KL divergence is also meaningful from an auditory point of view as the log term represents the difference in perceived loudness as approximated by Weber's law \cite{JesteadtWG77_LogFrequencyLoudnessStudy_JASA}:
\begin{align*}
\log(\spec\ind{m,n}/p_m) - & \log((\apspec)\ind{m,n}/p_m) = \\
& \qquad\qquad \log(\spec\ind{m,n}/(\apspec)\ind{m,n}),
\end{align*}
where $p_m$ is a frequency dependent constant related to the threshold of hearing.
Since our divergence is only defined for positive $\act$, we add the term
\[\objf_2(\act) := \charfunc_{\R_{\ge 0}^{K\times L \times N}}(\act)\]
to enforce non-negativity of $\act$, where $\charfunc_{S}$ is the \emph{characteristic function} of some set $S$, i.e. $\charfunc_{S}(\act)=\infty$ if $\act\notin S$ and zero otherwise.
Thus our current objective function becomes
\[\tobjf_b(\act) := \objf_1(\apspec) + \objf_2(\act).\]

To illustrate the effects resulting from the use of various cost and regularizer terms we employ in Fig.~\ref{fig:IlluRegul} a recording of bars three and four of Burgm{\"u}ller's Opus 100 Etude 2, see Fig.~\ref{fig:IlluRegul}a. Subfigures b-g show several $\act$ obtained by minimizing different objective functions. For a clearer visual representation of the order-3 tensor $\act$, we `flattened' $\act$ by stacking the slices $\act(k,:,:)$ vertically on top of each other to obtain a matrix representation of size $KL \times N$, parts of which are shown in Fig.~\ref{fig:IlluRegul}b-g.
For a semantically meaningful result, we expect to find for each note a diagonal line in $\act$, with the first template for the note being activated at the onset position followed by activations for the subsequent templates in the frames after the onset. Note that \figurename\ref{fig:IlluRegul} uses different scalings for the horizontal and vertical axis such that diagonal lines in $\act$ do not have $45$ degrees.

Looking at \figurename\ref{fig:IlluRegul}b, we see that after minimizing $\tobjf_b$ most activations indeed correspond to templates for the notes used in the example, compare \figurename\ref{fig:IlluRegul}a. However, a semantically meaningful diagonal structure hardly exists, with activation energy for a note being spread across different templates for a note (see bottom marker in \figurename\ref{fig:IlluRegul}b). Further, since we only have one spectral pattern corresponding to one specific playing style for each note in our dictionary tensor, the various patterns played in the actual recording somewhat differ from the ones in the dictionary. Therefore, some energy in $\spec$ corresponding to a specific note is not 'explained' by the corresponding pattern which leads to additional spurious activations (see upper marker in \figurename\ref{fig:IlluRegul}b). Overall, a clear discrimination between actual notes and estimation errors would be difficult using this $\act$.

\subsection{Encouraging Sparsity}
\label{subsec:sparsity}

To obtain fewer and more meaningful activation patterns, a typical approach is to encourage sparsity in $\act$ by adding its $\ell^1$ norm to the objective \cite{AbdallahP06_TranscriptionSparseCoding_TNN}:
\[\tobjf_c(\act) = \tobjf_b(\act) + \objf_3(\act),\]
where $\objf_3 := \lambda_1 \norm{\cdot}_1$ and $\lambda_1 \ge 0$ is a parameter balancing the importance of the sparsity and the remaining terms.
From a probabilistic point of view, the use of the $\ell^1$ norm is equivalent to assuming that the activities $\act$ we will encounter in real recordings are distributed according to a Laplace distribution \cite{ChenDS98_BasisPursuit_SIAMJSC,AbdallahP06_TranscriptionSparseCoding_TNN}. The Laplace can be interpreted as a variant of the Normal distribution that contains an absolute instead of a squared difference from its mean, which makes higher and sharper peaks in $\act$ more likely compared to the Normal distribution. While we could follow this probabilistic interpretation, we rather choose an optimization point of view as not all of our regularizers will have a straightforward probabilistic interpretation -- this will lead us to the well-known concept of \emph{soft thresholding} in Section~\ref{sec:parameterEst}.

In \figurename\ref{fig:IlluRegul}c we see the result of minimizing $\tobjf_c$ with respect to $\act$. The sparsifying and energy focusing effect of the $\ell^1$ term are clearly visible. A semantically meaningful diagonal activation structure, however, can still not be found but rather horizontal lines (bottom marker). The underlying reason is that we did not normalize our spectral templates as in other sparse coding method.
Due to the energy decay in piano sounds, the early templates in a pattern contain more energy than the subsequent ones. Therefore, we can often further minimize $\tobjf_c$ by activating a wrong template with less activation intensity, rather than using the correct template with more intensity. As a result something counterintuitive happens: increasing $\lambda_1$ can lead to more random, spurious activations because the use of the wrong templates leads to unexplained residual energy which is then modeled using other patterns (upper marker).

\subsection{Encouraging a Temporally Meaningful Template Order}

To counter these negative effects of the $\ell_1$ term, we next introduce a regularizer that enhances diagonal structures. To this end, we define for $(k,\ell,n) \in [1:K]\times [1:L-1] \times [1:N-1]$
\begin{align*}
\tdvop[\act](k,\ell,n) &:= \act(k,\ell,n) - \act(k,\ell+1,n+1),\\
\objf_4 &:= \lambda_2 \norm{\cdot}_1,\\
\tobjf_d(\act) & := \tobjf_c(\act) + \objf_4(\tdvop[\act])
\end{align*}
where $\lambda_2 \ge 0$ is another balancing parameter.
The $\tdvop$ operator is essentially a simple high-pass filter, which suppresses in combination with $\norm{\cdot}_1$ oscillations and unnecessary changes along the diagonals of $\act$.
Similar to temporal continuity constraints as used in NMF \cite{Virtanen07_MonauralSoundSourceSeparation_TASLP},
this approach corresponds to an anisotropic version of total variation image denoising, a well-studied problem in computer vision \cite{ChanEPY05_DevsTotalVariation_MMCV} where the goal is to remove various types of noise while preserving edges -- a property useful for us to account for the energy drop at offsets. 
Due to this similarity, we refer to this regularizer as \emph{total diagonal variation (TDV)}.

The results of minimizing $\tobjf_d(\act)$ with respect to $\act$ are shown in \figurename\ref{fig:IlluRegul}d. As we can see the TDV regularizer can be used to effectively attenuate not only semantically not meaningful horizontal and vertical  activation structures in $\act$ but also single spurious activations. Further, while the use of unnormalized templates had detrimental effects on the $\ell^1$ term, it increases the usefulness of the TDV regularizer drastically. In particular, as the energy decay of the piano sound is already accounted for in $\dict$, we expect only marginal change across a given diagonal in $\act$. This way, we can set a high value for $\lambda_2$ for the TDV term which suppresses many semantically meaningless activations without leading to strong negative effects from a modeling point of view. 

From a numerical point of view, it is important to note that $\tobjf_d$ is convex in $\act$. To see this, it is useful to recall which operations preserve convexity \cite{BoydV04_ConvexOptimization_Book}. Applying these rules, we see that $\distsmall$ is strictly convex in $b$, and thus $\dist$ applying $\distsmall$ in a sum to the entries in $\apspec$ is strictly convex in $\apspec$. Since $\apspec$ is linear in $\act$, $\dist(\spec,\apspec)$ is convex in $\act$. Further, $\charfunc_M$ is convex if $M$ is a convex set which is the case for $\R_{\ge 0}^{K\times L \times N}$. Finally, since $\norm{\act}_1$ is convex and  $\tdvop[\act]$ is linear in $\act$, we see that all terms in $\tobjf_d$ are convex in $\act$.
This convexity guarantees that our algorithms cannot get stuck in poor local minima of $\tobjf_d$, which leads to meaningful results even without a  good initialization for $\act$. 
On the downside, using just convex terms one is often limited to semantically less expressive regularizers. Therefore, as a general concept, we will use  $\tobjf_d$ primarily during a first algorithmic stage to robustly obtain a reasonable initialization for $\act$. Then, we add more expressive, non-convex regularizers that would often lead to useless results on their own -- in combination with this initialization, however, they can be used to further refine an already reasonable estimate of $\act$. 

\begin{figure}
\centering
\includegraphics[width=7.5cm]{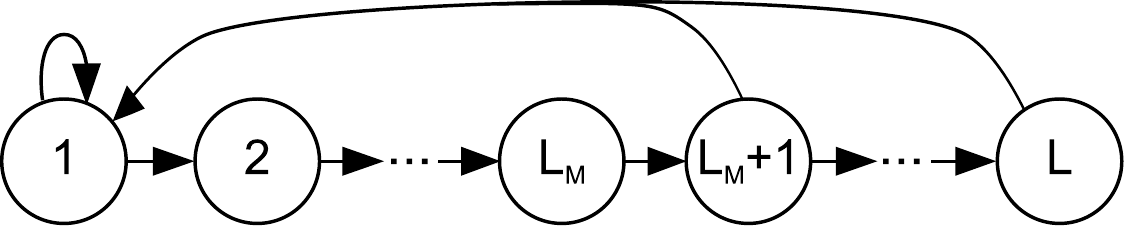}
\caption{Graphical model describing which transitions between templates for a specific piano key $k$ are encouraged by the regularizer term $\charfunc_\markov$.}
\label{fig:graphModelSoundProgression}
\end{figure}

\subsection{Constraining the Concurrency of Templates}
\label{sec:MarkovStateReg}

To see remaining issues, note that in \figurename\ref{fig:IlluRegul}d for a single piano key $k$ several templates can be active at the same time in parallel diagonal lines (see markers), which is semantically not meaningful and can lead to estimation errors. 
To mitigate such issues, we now develop a regularizer which encourages activations in $\act$ to follow transition rules for the templates described by the graphical model depicted in \figurename\ref{fig:graphModelSoundProgression}, i.e. we integrate the concepts behind the FS-HMM and its variants \cite{OzerovFC09_FSHMM_WASPAA} (or heuristic approaches as in \cite{DriedgerPM15_AudioMosaicingNMF_ISMIR}).
While one of our contributions is a more robust parameter estimation process for these models (to be described in Section~\ref{sec:parameterEst}),
we focus for now on the conceptual differences from a modeling point of view. In particular, instead of using a fully connected ergodic process, we employ, from a probabilistic point of view, a structured Bakis-type Markov process.
Using such a process, we can express that we expect a specific sequence of templates after an onset, including a minimum duration for the entire sequence.

More precisely, for a given $k$, the model expresses that if we want to use the $\ell$-th template in frame $n$, we need to use template $\ell-1$ in frame $n-1$, with an exception only for the first template.
Further, to use template $\ell$ for $2$\,$\le$\,$\ell$\,$<$\,$L_M$ in a given frame we must use templates $\ell+1,\ldots,L_M$ in the subsequent frames, where $L_M$ is referred to as the \emph{minimum note length} in frames.
To realize this model as a regularizer, we employ the characteristic function $\charfunc$ in combination with a specific set $\markov \subset \R_{\R\ge 0}^{K\times L \times N}$, and $\act \in \markov$ if for each $k$ the corresponding $\act(k,:,:)$ encodes, in a specific way, a valid state sequence for our graphical model. A simple encoding could be to define that state $\ell$ is considered active in frame $n$ if $\act(k,\ell,n)>0$ and the remaining entries in $\act(k,:,n)$ are zero. However, since our time-frequency representation $\spec$ will use overlapped windows, this would be too restrictive. In particular, each part of the signal is contained in several consecutive frames and therefore we need the capability in our model to activate consecutive templates together in a given frame.
Therefore, we define that $\act(k,:,n)$ encodes state $\ell$ if $\act(k,\tilde{\ell},n)\ge 0$ for all $\tilde{\ell} \in \{\ell,\ldots,\ell+\lceil w/s\rceil\}\cap\{1,\ldots,L\}$ and $\act(k,\tilde{\ell},n)= 0$ for all remaining $\tilde{\ell}$, where $w$ and $s$ are the window and step size in samples used to compute the spectrogram $\spec$. Due to its resemblance of Markov models, we refer to $\charfunc_\markov(\act)$ as the \emph{Markov-State (MS) regularizer}.
The results of initializing $\act$ using $\tobjf_d$ and then refining it with
\[\tobjf_e(\act) := \tobjf_d(\act) + \objf_5(\act)\]
with $\objf_5 := \charfunc_\markov$ are shown in \figurename\ref{fig:IlluRegul}e. As we can see, including the MS term removes the concurrent activations highlighted in \figurename\ref{fig:IlluRegul}d and the structure described by the graphical model leads to semantically more meaningful activation patterns. 

\subsection{Differentiating between Estimation Noise and Note Events}
\label{sec:thresholding}

Inspecting \figurename\ref{fig:IlluRegul}e, however, reveals a further problem. In particular, while the sparsity and TDV terms used in $\tobjf_d$ led to a reduction of spurious activations some still remain, see markers. To remove these, we now make use of the example recording of a note played pianissimo (low intensity).
Here, the idea is to provide an intuitive way for the user to specify a threshold used to differentiate between estimation noise and actual notes, which is in contrast to many previous methods where the user needs to adjust this value and having to decide which setting works  best -- each time the recording level or other recording conditions change.
As an additional benefit of having this threshold before the optimization process starts, we can include this knowledge as part of the optimization process.
For the given low-intensity recording, we compute an activation tensor $\act_M$ using $\tobjf_e$ in a pre-processing step and define our threshold as $\minact := \max_{k,\ell,n}\act_M(k,\ell,n)$.
Everything below this threshold will be considered as noise.
However, instead of just thresholding $\act$ after the estimation as usually done,  we make the optimization process aware of this requirement and integrate it as an additional regularizer.
Thus the energy below the threshold is not truncated but can be explained within the model leading to clearer activations.
To this end, we use the characteristic function in combination with the set $\threshset := \big( [\minact,\infty) \cup \{0\} \big)^{K \times L \times N}$.
\figurename\ref{fig:IlluRegul}f shows that using $\tobjf_d$ to obtain an initialization followed by a refinement using $\tobjf_f(\act) := \tobjf_e(\act) + \objf_6(\act) $ with $\objf_6 := \charfunc_\threshset$ eliminates most estimation noise from $\act$.

Activations estimated using $\tobjf_f$ typically correspond closely to the notes being played and
onset positions can be estimated to a high accuracy.
The offset position, however, is not always well captured. This is an effect of using unnormalized templates with the sparsity and TDV regularizers, which sometimes shortens and sometimes extends notes to make up for estimation errors. This can often be individually corrected by adjusting the  parameters $\lambda_1$ and $\lambda_2$ -- however, the right value depends on the recording. In \figurename\ref{fig:IlluRegul}f we chose values for $\lambda_1$ and $\lambda_2$ leading to such an issue, see marker. 

\subsection{Encouraging Meaningful Long-Term Note Activity}
\label{subsec:BinMarkovStrictCoupl}

As a starting point to eliminate these effects, we can observe in \figurename\ref{fig:IlluRegul}f that the activation intensity for affected notes typically drops at the correct offset position (see marking). As a countermeasure we now disallow any change in activation value while a note is active, which we implement by changing the signal model: we replace $\act$ in Eq.\ref{eq:mainModel1} by the Hadamard product of two matrices: $\actB \odot \actG$. 
While $\act$ encoded both the activation intensity and length of the note, we split these responsibilities into these two matrices, which enables more direct constraints. 
In particular, we constrain $\actG$ in such a way that, for each diagonal in $\actG(k,:,:)$ for a given $k$, the entries are only allowed to use a single, shared value, i.e. a strict coupling of values across frames. Since note-lengths cannot be modeled in $\actG$, we multiply it pointwise with the binary matrix $\actB$ which is subject to constraints similar to our Markov-state regularizer (Section~\ref{sec:MarkovStateReg}). 

More precisely, to obtain estimates for $\actB$ and $\actG$ we minimize
\[\tobjf_g(\actB,\actG) := \dist(\spec,\apspecBG) + \charfunc_{\widetilde{\markov}}(\actB) + \charfunc_{\widetilde{\threshset}}(\actG),\]
where $\widetilde{\markov} := \markov \cap \{0,1\}^{K \times L \times N}$ and $\widetilde{\threshset} := \{A \in \threshset | \; \norm{\tdvop[A]}_1 = 0 \}$. This is a highly non-convex objective function, so we rely on a meaningful initialization to obtain useful results. Here, we use the $\act$ obtained via $\tobjf_f$ as follows. First, we set $\actB(k,\ell,n)=1$ if $\act(k,\ell,n) > \minact$, and zero otherwise. Second, we set $\actG(k,\ell,n)= \max \{\act(k,1,n-\ell+1),\act(k,2,n-\ell+2),\ldots,\act(k,L,n-\ell+L)\}$: the maximum over the $(n-\ell+1)$-th diagonal in $\act(k,:,:)$. The effect of using these \emph{binary Markov-state} and \emph{strict coupling regularizers} is shown \figurename\ref{fig:IlluRegul}g. In particular, the use of $\charfunc_{\widetilde{\threshset}}$ indeed leads to constant values on the diagonals of $\act := \actB \odot \actG$. These values are typically considerably higher than the incorrect values we observed before following an offset position in $\act$.
As a result, we found that the parameter estimation for $\actB$ typically switches from the value $1$ to $0$ on a diagonal around the correct offset, as otherwise the high values on the diagonal would lead to high error rates after the offset.
As a drawback, however, we sometimes observed that a note is shortened too much and the resulting residual energy can lead to new note activations. An example is highlighted in \figurename\ref{fig:IlluRegul}g. For this reason, it seems this last combination of regularizers is more useful for correcting note lengths rather than obtaining a full transcription. 

\section{Parameter Estimation using the Alternating Direction Method of Multipliers}
\label{sec:parameterEst}

So far, we focused on the design of meaningful objective functions and their effect on the parameter estimation. In this section, we develop robust algorithms to minimize %
them.
A major problem is that our functions contain non-differentiable (e.g. $\norm{\cdot}_1$), infinite (e.g. $\charfunc_{\R_{\ge 0}^{K\times L \times N}}$) as well as non-convex terms (e.g. $\charfunc_{\markov}$): many classical optimization methods are not stable under these properties.
In this context, the \emph{alternating direction method of multipliers (ADMM)} has sparked a lot of interest in recent years \cite{BoydPCPE11_ADMM_Book,CombettesP11_ProximalMethodsInSP_BookChap}. ADMM belongs to a class %
referred to as \emph{proximal algorithms}, whose importance Parikh and Boyd \cite{ParikhB13_ProxAlgor_FTO} describe as ``much like Newton's method is a standard tool for solving
unconstrained smooth optimization problems of modest size, proximal
algorithms can be viewed as an analogous tool for non-smooth, constrained,
large-scale, or distributed versions of these problems''. In general, ADMM is of  interest if the objective function comprises several terms that are difficult to minimize jointly but efficiently individually. In this respect, ADMM provides a scheme to split up the objective function, minimize the terms individually and still provide convergence guarantees for the entire objective. As a result it is not only useful for complex objective functions as in our case but also in big data scenarios, as ADMM's splitting and merging operations fit perfectly into distributed computing schemes like \emph{Map-Reduce}. For a comprehensive introduction, we refer to \cite{BoydPCPE11_ADMM_Book,CombettesP11_ProximalMethodsInSP_BookChap,ParikhB13_ProxAlgor_FTO}.

ADMM solves problems of the form 
\begin{align}
\label{eq:ADMM1}
\argmin_{x,z} \qquad & f(x) + g(z)\\
\text{subject to} \qquad & {B}x + {C}z = c \nonumber
\end{align}
with $x \in \R^N$,  $z \in \R^M$, $B \in \R^{P \times N}$, $C \in \R^{P \times M}$, $c \in \R^P$.
Solutions for problem (\ref{eq:ADMM1}) are identical to the ones for the following problem (under some assumptions \cite{BoydV04_ConvexOptimization_Book}) 
\begin{equation}
\argmax_{\beta} \inf_{x,z} L_\rho(x,z,\beta),
\label{eq:ADMMAugmLagr}
\end{equation}
\vspace{-0.4cm}
\begin{align*}
L_\rho(x,z,\beta) := f(x) + g(z) & + \langle \beta , Bx + Cz - c\rangle \\
& + (\rho/2) \norm{Bx + Cz - c}^2_2
\end{align*}
where $L_\rho$ is referred to as the \emph{augmented Lagrangian} for problem (\ref{eq:ADMM1}) with \emph{penalty parameter} $\rho>0$ and \emph{dual variable} $\beta \in \R^P$ (see e.g. \cite{BoydV04_ConvexOptimization_Book,BoydPCPE11_ADMM_Book}). A classical method to iteratively solve (\ref{eq:ADMMAugmLagr}) is to minimize $L_\rho$ w.r.t. $x$ and $z$, followed by a maximization over $\beta$. Convergence guarantees only hold if the minimization is computed jointly in $x$ and $z$, which in practice is often difficult. The main extension in ADMM is to split the minimization into two steps:
\begin{align}
\label{eq:ADMMAlgor}
x^{k+1} & := \argmin_x L_\rho(x,z^k,\beta^k)\\
z^{k+1} & := \argmin_z L_\rho(x^{k+1},z,\beta^k) \nonumber\\
\beta^{k+1} & := \beta^{k} + \rho (Bx^{k+1} + Cz^{k+1} - c )\nonumber,
\end{align}
where $k$ is the iteration number.
Note that the update of $\beta$ is a gradient ascent on $L_\rho(x^{k+1},z^{k+1},\cdot)$ with stepsize $\rho$. Eckstein and Bertsekas \cite{EcksteinB92_ProofADMM_MP} showed that ADMM is equivalent to a firmly non-expansive operator, which enables the application of  theorems %
closely related to the well-known Banach fixed-point theorem.
As a consequence, despite the split, ADMM converges for any $\rho$\,$>$\,$0$ under relatively mild conditions. In particular, there is no need for $f$ or $g$ to be differentiable or strictly convex or even finite (i.e. they can assume the value $\infty$) -- they only need to be convex, as well as closed and proper. Further, the $x$ and $z$ updates do not even have to be exact but can be approximate to some degree. Since the initial proof \cite{EcksteinB92_ProofADMM_MP}, convergence results were greatly extended, see \cite{BoydPCPE11_ADMM_Book,CombettesP11_ProximalMethodsInSP_BookChap,ParikhB13_ProxAlgor_FTO} and references therein.

\subsection{Consensus Form ADMM}
\label{sec:ConsADMM}

Though not obvious at first, the splitting of the minimization step in ADMM has considerable consequences and enables us to divide our objective functions into their constituent terms, which are much easier to minimize individually.
First, we identify our problem to be of the form used in Eq.~(\ref{eq:ADMM1}). To this end, we will treat $\act$ as an element of a vector space and do not differentiate between $\R^{M\times L\times K}$ and $\R^{MLK}$.
Our objective functions $\tobjf_a$ to $\tobjf_f$ have the following form
\begin{equation}
\argmin_{\act} \sum^{I}_{i=1} f_i(C_i \act),
\label{eq:probOursOrg}
\end{equation}
where $C_i$ are linear operators corresponding to the dictionary pattern operator $\dict$, the TDV operator $\tdvop$ or simply the identity operator.
Note that, with $\act$ interpreted as an element of $\R^{MLK}$, each $C_i$ could be represented by a matrix in $\R^{M_i\times MLK}$ for some $M_i>0$. 
Problem (\ref{eq:probOursOrg}) is equivalent to:
\begin{align}
\label{eq:probOursOrg2}
\argmin_{x_1,\ldots,x_I,\act}  \qquad & \sum^{I}_{i=1} f_i(x_i)\\
\text{subject to} \qquad & \forall i \in \{1,\ldots,I\}: x_i= C_i \act,  \nonumber
\end{align}
where $x_i \in \R^{M_i}$.
This form is often referred to as the \emph{consensus form} of problem~(\ref{eq:probOursOrg}) since all local variables $x_i$ have to agree on a common solution enforced by the constraints \cite{NedicO10_DistributedOptimization_COSPC}. In this context, $\act$ is called the \emph{central collector}.
By setting $X := (x_1,\ldots,x_I)$, $f(X) := \sum^{I}_{i=1} f_i(x_i)$, $\mathcal{\act} := \{ \copyact := (\act_1,\ldots,\act_I) \in \R^{MLKI}| \act_1=\ldots=\act_I\}$ and $g = \charfunc_\mathcal{\act}$, we see that problem~$(\ref{eq:probOursOrg2})$ is equivalent to:
\begin{align}
\label{eq:ADMM2}
\argmin_{X,\copyact} \qquad &f(X) + g(\copyact)\\
\text{subject to} \qquad & X - C \copyact = 0 \nonumber
\end{align}
where $C$ is a block-diagonal matrix 
\begin{equation}
C := \begin{pmatrix}
C_1 & & \\
  & \ddots & \\
  & & C_I\\
\end{pmatrix}.
\label{eq:defB}
\end{equation}
Here, the characteristic function $\charfunc_\mathcal{\act}$ ensures that the copies of $\act$ in $\copyact$ are identical.
Problem~(\ref{eq:ADMM2}) clearly adheres to the form shown in Eq.~(\ref{eq:ADMM1}) and thus we can apply ADMM.

To see advantages of this form compared to (\ref{eq:probOursOrg}), we first note that the augmented Lagrangian $L_\rho$ for problem~(\ref{eq:ADMM2}) is \emph{separable} by construction
in each element of $\{x_1,\ldots,x_I\}$, i.e. $L_\rho(\cdot, Z^k, \beta^k)$ partitions its input and processes each disjunct subset independently. 
More precisely, the augmented Lagrangian for problem (\ref{eq:ADMM2}) can be written as
\[L_\rho(X,\copyact,\beta) := \sum^{I}_{i=1}  f_i(x_i) + \langle \beta_i , x_i - C_i \act_i\rangle + \frac{\rho}{2} \norm{x_i - C_i \act_i}^2_2\]
\[ + \charfunc_\mathcal{\act}(\act_1,\ldots,\act_I) \qquad\qquad\;\]
where the partition of $\beta = (\beta_1,\ldots,\beta_I)$ is defined analogously to the partition of $X$. 
As we can see, a specific $x_i$ appears only in exactly three terms, and every other term is independent of $x_i$. 
Thus, we can update each $x_i$ individually, which enables us to develop specialized, efficient minimizers for each term -- which can even run in parallel. More precisely, the $X$ update for our problem is equivalent to (for all $i \in \{1,\ldots,I\}$):
\begin{equation}
x^{k+1}_i = \argmin_{x_i} f_i(x_i) + \langle \beta^k_i , x_i - C_i \act^k_i\rangle +\frac{\rho}{2} \norm{x_i -  C_i  \act^k_i}^2_2
\label{eq:separatedTermsUpdate}
\end{equation}
We develop specific solvers for problem~(\ref{eq:separatedTermsUpdate}) for $f_1$ to $f_6$ in Section~\ref{sec:minIndRegs}.

\subsection{Linearized ADMM}

Once all $x_i$ are computed, we can update $\copyact$, which corresponds to the $z$ update in Eq.~(\ref{eq:ADMMAlgor}), i.e. we need to solve:
\begin{equation}
\argmin_{\copyact} \charfunc_\mathcal{\act}(\copyact) + \langle \beta^k , X^{k+1} - C\copyact \rangle + \frac{\rho}{2} \norm{X^{k+1} -C\copyact}^2_2.
\label{eq:actupdate1}
\end{equation}
We can pull the inner product into the norm to obtain the equivalent \emph{proximal form} (leaving out terms independent of $\copyact$)
\[\argmin_{\copyact} \charfunc_\mathcal{\act}(\copyact) + \frac{\rho}{2} \norm{C\copyact - (X^{k+1} + (1/\rho) \beta^k)}^2_2.\]
We will use this form repeatedly in the next section.
A problem of this form can be identified as a quadratic programming problem with linear equality constraints, for which specific solvers exist \cite{BoydV04_ConvexOptimization_Book}.
Such methods, however, compute exact solutions and are computationally too expensive for our purposes.

As discussed in \cite{ZhangBO11_LinearizedADMM_JSC}, problem~(\ref{eq:actupdate1}) is easily solved if $C$ would be the identity matrix -- in this case we only need to compute an orthogonal projection of $X^{k+1} + (1/\rho) \beta^k$ onto $\mathcal{\act}$ which turns out to be straightforward. Following this idea, we now try to extract $C$ out of the norm in (\ref{eq:actupdate1}). 
To this end, we use here a concept similar to the one presented in \cite{ZhangBO11_LinearizedADMM_JSC}, referred to as \emph{Linearized ADMM}, and thus our approach can be considered as a combined \emph{Linearized Consensus ADMM}. More precisely, we replace the term $(\rho/2) \norm{X^{k+1} -C\copyact}^2_2$ with
\[\rho \langle C^\top C \copyact^k - C^\top \xlinadmm, \copyact \rangle + (\mu/2) \norm{\copyact-\copyact^k}^2_2,\]
for some $\mu\geq \rho \norm{C}^2$ where $\norm{\cdot}$ is the spectral norm, and $\copyact^k$ and $X^{k+1}$ denote estimates computed in the $k$-th and $k+1$-th iteration, respectively. This change can be interpreted as adding additional regularizers to the augmented Lagrangian and is sometimes referred to as the \emph{inexact Uzawa} approach -- see \cite{ZhangBO11_LinearizedADMM_JSC} for a more in-depth discussion. With this change, the $\copyact$ update becomes (again pulling all inner products into the norm as above)
\[\argmin_{\copyact} \charfunc_\mathcal{\act}(\copyact) + \frac{\rho}{2} \big\| \copyact - (\copyact^k - \frac{\rho}{\mu} C^\top (C\copyact^k - \xlinadmm - \frac{1}{\rho}\beta^k) ) \big\|^2_2.\]
With our variable $\copyact$ freed of $C$, the solution simply becomes the orthogonal projection $\pi$ of $\copyact^k - \frac{\rho}{\mu} C^\top (C\copyact^k - \xlinadmm - \frac{1}{\rho}\beta^k)$ onto the set $\mathcal{\act}$:
\[
\copyact^{k+1} = \pi_\mathcal{\act}(\copyact^k - \frac{\rho}{\mu} C^\top (C\copyact^k - \xlinadmm - \frac{1}{\rho}\beta^k)),
\]
which due to the definition of $\mathcal{\act}$
corresponds to simply taking the average over its components, i.e. with
\[\act^{k+1} := \frac{1}{I}  \sum^I_{i=1} \act^k_i - \frac{\rho}{\mu} C^\top_i ( C_i \act^k_i - x^{k+1}_i-(1/\rho) \beta^k_i)\]
the updated variable is then $\copyact^{k+1} := (\act^{k+1},\ldots,\act^{k+1})$. Note that $\act^{k+1}=\act_1^{k+1}=\ldots=\act_I^{k+1}$ is guaranteed, which allows us to leave out the index $i$ on $\act^k$ in the following and simplify the update even further:
\begin{equation}
\act^{k+1} := \frac{1}{I}  \sum^I_{i=1} \act^k - \frac{\rho}{\mu} C^\top_i ( C_i \act^k - x^{k+1}_i-(1/\rho) \beta^k_i)
\label{eq:actUpdate}
\end{equation}
As shown in \cite{ZhangBO11_LinearizedADMM_JSC} convergence results also hold for this ADMM variant.

As a drawback of this modification, however, we found that the condition $\mu\geq \rho \norm{C}^2$ required to prove convergence is too restrictive in practice and the resulting method might require an excessive amount of iterations to converge.
As an alternative, we developed a second update method for $\copyact$, which is based on the introduction of additional slack variables.
With this method, we achieved an improved convergence rate compared to linearized ADMM.
Existing convergence proofs, however, do not hold for this approach anymore. Due to space constraints this variant is presented in an external annex \cite{Ewert16_AppendixADFPT_TechReport}. 

Finally, to complete our ADMM framework for problem~(\ref{eq:ADMM2}), we give the update rules for $\beta$. Here, we can exploit the block structure of $C$ and express the update more compactly for all $i \in \{1,\ldots,I\}$ as
\begin{equation}
\beta^{k+1}_i := \beta^k_i + \rho (x^{k+1}_i - C_i \act^{k+1}).
\label{eq:betaUpdate}
\end{equation}

\section{Minimizing the Individual Terms}
\label{sec:minIndRegs}

With the general framework in place, we now develop methods for minimizing the individual terms in our objective function, i.e. implement Eq.~\ref{eq:separatedTermsUpdate} for each regularizer or data fidelity term. For a lack of space we omit most proofs but point to relevant work for those terms that had been introduced previously in similar forms.

\subsection{Kullback-Leibler Data Fidelity Term}

We start with the Kullback-Leibler term $\dist$, for which Eq.~\ref{eq:separatedTermsUpdate} has the following form:
\[x^{k+1}_1 := \argmin_{x_1} \dist(\spec,x_1) + \langle \beta^k_1, x_1 -  \apspecvar{\act^k}\rangle + \frac{\rho}{2}\norm{x_1 -  \apspecvar{\act^k}}^2_2.\]
Without a linear transform in the arguments of $\dist$ (result of our construction), the calculation becomes straightforward and $x^{k+1}_1$ can be found analytically:
The gradient of the right hand side is $\beta^k_1 - \frac{\spec}{x_1} + \rho (x_1 -  \apspecvar{\act^k}) + 1$,
where the division is element-wise and $1$ is the all-one vector.
To find the minimizing argument, we set the gradient to zero and obtain
\begin{solver}
x^{k+1}_1 = \frac{\rho \apspecvar{\act^k} - \beta^k_1 - 1 + \sqrt{(\rho \apspecvar{\act^k} - \beta^k_1 - 1)^2 + 4 \rho \spec}}{2\rho},
\label{eq:updateKl}
\end{solver}
where the square and square root are again element-wise. %

\subsection{Non-negativity Term}

The non-negativity term does not include a linear transformation, i.e. $C_2$ is the identity matrix, and problem~(\ref{eq:separatedTermsUpdate}) has the following form
\[x^{k+1}_2 := \argmin_{x_2} \charfunc_{\R^{K \times L \times N}_{\ge 0}}(x_2) + \langle \beta^k_2, x_2 -  \act^k\rangle + \frac{\rho}{2}\norm{x_2 -  \act^k}^2_2,\]
which in proximal form is
\[x^{k+1}_2 := \argmin_{x_2} \charfunc_{\R^{K \times L \times N}_{\ge 0}}(x_2) + \frac{\rho}{2}\norm{x_2 -  (\act^k- \frac{1}{\rho} \beta^{k}_2)}^2_2.\]
As already seen above, the solution here is an orthogonal projection of $\act^k- \frac{1}{\rho} \beta^{k}_2$ onto the set $\R^{K \times L \times N}_{\ge 0}$, i.e. the solution is a half-wave rectifier \cite{BoydPCPE11_ADMM_Book}:
\begin{solver}
x^{k+1}_2 := \pi_{\R^{K \times L \times N}_{\ge 0}}(\act^k- \frac{1}{\rho} \beta^{k}_2) = \max(\act^k- \frac{1}{\rho} \beta^{k}_2,0),
\label{eq:updateNonNeg}
\end{solver}
where $\max$ is entrywise and $0$ is the all-zero matrix.

\subsection{LASSO and Total Diagonal Variation Terms}

Problem~(\ref{eq:separatedTermsUpdate}) in proximal form for the LASSO or $\ell_1$ term is
\[x^{k+1}_3 := \argmin_{x_3} \norm{x_3}_1 + \frac{\rho}{2 \lambda_1}\norm{x_3 -  ( \act^k - \frac{1}{\rho}\beta^k_3)}^2_2.\]
A problem in computing $x^{k+1}_3$ is that $\norm{\cdot}_1$ is not differentiable everywhere and thus we cannot simply set the gradient to zero as before. However, similar strategies are possible when we use the sub-derivative instead, which is set valued and, compared to the derivative, contains all possible linearizations of the function that do not locally exceed the function value \cite{Rockafellar70_ConvexAnalysis_PMS}. For example, the sub-derivative for the absolute value function is either $\{-1\}$ or $\{1\}$ everywhere, except for zero where it is the closed interval $[-1,1]$. In this context, analogue to the derivative, a minimum of a convex function is characterized by its sub-derivative containing the value zero. The solution to the above problem derived this way is commonly referred to as \emph{soft-thresholding} \cite{CombettesP11_ProximalMethodsInSP_BookChap}: 
\begin{solver}
x^{k+1}_3 := \sign(v) \max(|v| - \frac{\lambda_1}{\rho},0),
\label{eq:updateSparse}
\end{solver}
where $v:=\act^k - \frac{1}{\rho}\beta^k_3$ and all operations are element-wise. 

The main difference between the LASSO and our new Total Diagonal Variation term is the application of the $\tdvop$ operator. This operator is linear in its argument and thus could be represented by a matrix, i.e. it takes the role of the matrix $C_4$ in the last section.
Due to our construction the $\tdvop$ operator is not applied to $x_4$ (but to the central collector $\act$), which again turns out to simplify our minimization problem considerably: The solution for the TDV term is almost identical to the LASSO term, with
\begin{solver}
x^{k+1}_4 := \sign(v) \max(|v| - \frac{\lambda_2}{\rho},0)
\label{eq:updateTDV}
\end{solver}
and %
$v:=\tdvop \act^k - \frac{1}{\rho}\beta^k_4$.

\subsection{Markov-State Regularizer}

For our new Markov-State regularizer we have to solve the following problem (proximal form)
\begin{align}
\label{eq:probMarkovState}
x^{k+1}_5 & := \argmin_{x_5} \charfunc_{\markov}(x_5) + \frac{\rho}{2}\norm{x_5 -  (\act^k- \frac{1}{\rho} \beta^{k}_5)}^2_2\\
& = \pi_{\markov}(\act^k- \frac{1}{\rho} \beta^{k}_5) \nonumber
\end{align}
To compute this orthogonal projection, we can use dynamic programming. Setting $v := \act^k- \frac{1}{\rho} \beta^{k}_5 \in \R^{K \times L \times N}_{\ge 0}$, we define for each $k$ a cost matrix $C_k \in \R^{L \times N}$
\begin{equation}C_k(\ell,n) = \sum_{\mathclap{\quad\quad \tilde{\ell} \in \{1,\ldots,L\} \setminus \{\ell,\ldots,\ell+\lceil w/s\rceil\}}}v(k,\tilde{\ell},n)^2.
\label{eq:costMatrixMS}
\end{equation}
This is the error (w.r.t. the squared Euclidean norm in Eq.~\ref{eq:probMarkovState}) we make in frame~$n$ if we encode state~$\ell$ in that frame for key~$k$  (compare also Section~\ref{sec:MarkovStateReg}).
Using $C_k$ and dynamic programming, we can find the state sequence $\ell^k_{1},\ldots,\ell^k_{N}$ minimizing the total cost $\sum_n C_k\ind{\ell^k_{n},n}$ among 
all sequences valid under the graphical model shown in \figurename~\ref{fig:graphModelSoundProgression}.
To this end, we recursively define an accumulated cost matrix $D_k \in \R_{\ge 0}^{L \times N}$ and a step matrix $E_k \in \{1,\ldots,L\}^{L \times N}$ as
\begin{equation}
D_k\ind{\ell,n} := C_k(\ell,n) + 
\begin{cases}
D_k\ind{\ell-1,n-1}, & \ell>1\\
\displaystyle\min_{\tilde{\ell} \in \mathcal{S}}(D_k\ind{\tilde{l},n-1}) & \ell=1
\end{cases}
\label{eq:dynProg}
\end{equation}
\begin{equation}
E_k\ind{\ell,n} := \begin{cases}
\ell-1, & \ell>1\\
\displaystyle\argmin_{\tilde{\ell} \in \mathcal{S}}(D_k\ind{\tilde{\ell},n-1}) & \ell=1
\end{cases}
\label{eq:dynProg2}
\end{equation}
where $\mathcal{S} := \{1,L_M,L_M +1,\ldots,L\}$ is the set of states that allow a return to the first state and $D_k\ind{\ell,1} := C_k(\ell,1)$ for all $\ell$.
We start by setting $\ell^k_N = \argmin_\ell D_k(\ell,N)$ and set $\ell^k_n = E_k\ind{\ell^k_{n+1},n+1}$ for $n \in \{1,\ldots,N-1\}$. Note that the definition of $\mathcal{S}$ only allows state transition that are valid according to our graphical model (including the minimum note duration $L_M$). Having the state sequence, $x^{k+1}_5$ is
\begin{solver}
x^{k+1}_5(k,\ell,n) :=
\begin{cases}
v(k,\ell,n), & \ell \in \{\ell^k_n,\ldots,\ell^k_n+\lceil w/s\rceil\} \\
0 & \text{otherwise}
\end{cases}
\label{eq:updateMS}
\end{solver}

Due to the simplicity of our graphical model, implementations of Eqns.~\ref{eq:costMatrixMS} to \ref{eq:dynProg2} can be highly efficient. First, all computations can be parallelized over $k$. Second, $C_k$ can be computed by one element-wise multiplication of $v$ followed by one matrix multiplication (for the sum). Third, for $D_k$ and $E_k$ we only need to compute the minimizer for $\ell=1$ and the other entries can be computed independently given the results for the previous frame. Therefore, the computation can be almost perfectly parallelized over $\ell$ as well. Overall, we found computing the solution for this regularizer to be only marginally slower compared to the other regularizers.

\subsection{Thresholding Set Term}

\begin{figure}
\vspace{-0.3cm}  %
\begin{algorithm}[H]
 \caption{ADMM for Minimizing $\tobjf_b$ to $\tobjf_f$}\label{Alg:MinBToF}
 \begin{algorithmic}[1]
\State \textbf{Initialization:}
\State \hspace{0.5cm} Set penalty $\rho$ to a positive value.
\State \hspace{0.5cm} Set $I$ to the number of terms in the objective function.
\State \hspace{0.5cm} Initialize $\act$ with random values.
\State \hspace{0.5cm} \textbf{For $i=1$ to $I$:}
\State \hspace{1cm} Set $x_i$ to $C_i \act$.
\State \hspace{1cm} Set $\beta_i$ to $0$. \Comment{all-zero vector}
\State \textbf{Repeat Until Convergence:}
\State \hspace{0.5cm} \textbf{For $i=1$ to $I$:}
\State \hspace{1cm} Update $x_i$ using Eq.~$S_i$.
\State \hspace{0.5cm} Update $\act$ using Eq.~\ref{eq:actUpdate}.
\State \hspace{0.5cm} \textbf{For $i=1$ to $I$:}
\State \hspace{1cm} Update $\beta_i$ using Eq.~\ref{eq:betaUpdate}.
 \end{algorithmic}
 \end{algorithm}
\end{figure}

For the thresholding set term, we need to solve
\begin{align}
x^{k+1}_6 & := \argmin_{x_6} \charfunc_{\threshset}(x_6) + \frac{\rho}{2}\norm{x_6 -  (\act^k- \frac{1}{\rho} \beta^{k}_6)}^2_2\\
& = \pi_{\threshset}(\act^k- \frac{1}{\rho} \beta^{k}_6), \nonumber
\end{align}
where $\threshset := \big( [\minact,\infty) \cup \{0\} \big)^{K \times L \times N}$ (repeated from Section~\ref{sec:thresholding}). Using again $v$ as a shorthand, setting $v := \act^k- \frac{1}{\rho} \beta^{k}_6$, the solution is
\begin{solver}
x^{k+1}_6(k,\ell,n) := \begin{cases}
v(k,\ell,n), & \minact < v(k,\ell,n) \\
\minact & \frac{\minact}{2} \leq v(k,\ell,n) \leq \minact\\
0 & \text{otherwise.}
\end{cases}
\label{eq:updateThresh}
\end{solver}
This completes our method for minimizing $\tobjf_b$ to $\tobjf_f$. Algorithm~\ref{Alg:MinBToF} summarizes the individual steps in pseudo code.  Further improvements are described in the external annex \cite{Ewert16_AppendixADFPT_TechReport}, where an alternative update rule for $\act$ is derived, which converged quicker in practice than the linear ADMM based solution presented here, and presents an adaptive scheme, which updates the penalty parameter $\rho$ after each iteration to increase the convergence speed (using similar ideas as presented in \cite{BoydPCPE11_ADMM_Book}, where ADMM is discussed as a primal-dual algorithm \cite{Rockafellar70_ConvexAnalysis_PMS}, which leads to an effective heuristic for adjusting $\rho$ in such a way that the primal and dual residual are balanced). Additionally, a Matlab implementation is available online\footnote{\url{https://code.soundsoftware.ac.uk/projects/adpt}}.

\subsection{Binary Markov-State and Strict Coupling Regularizer}
\label{sec:NoteLengthModel}

For the binary Markov-state and strict coupling regularizers we do not use ADMM but minimize the augmented Lagrangian directly. We do this for two reasons. First, it shows that the augmented Lagrangian is often also useful without ADMM. Second, since our objective function $\tobjf_g$ is non-convex, the convergence guarantees provided by ADMM do not hold anymore.
Due to space constraints, we only present a summary of the iterative updates here, with details on the derivation in an external annex \cite{Ewert16_AppendixADFPT_TechReport}.

In general, we use the augmented Lagrangian for the following problem, which is equivalent to our objective function $\tobjf_g$,
\begin{align}
\label{eq:BMSSC_prob}
\argmin  \qquad &  \dist(\spec,X_1) + \charfunc_{\widetilde{\markov}}(X_5) + \charfunc_{\widetilde{\threshset}}(X_6)\\
\text{subject to} \qquad & X_1 = \apspecvar{X_2}, \quad X_2 = X_3 \odot X_4 \nonumber \\
& X_3 = X_5, \quad X_4 = X_6 \nonumber
\end{align}
The additional slack variables in the equality constraints are introduced to simplify the decoupling of the actual variables. Minimizing for $X_1$ to $X_6$ individually, we obtain the following update rules
\begin{flalign*}X_1 & = \frac{\rho \apspecvar{X_2} - \beta_{X_1} - 1 + \sqrt{(\rho \apspecvar{X_2} - \beta_{X_1} - 1)^2 + 4 \rho \spec}}{2\rho} & \end{flalign*}
\[X_2 = (\dict^\top \dict + I)^{-1} (\dict^\top X_1
+  X_3 \odot X_4 + (1/\rho) (\dict^\top \beta_{X_1} - \beta_{X_2})   )\]
\[X_3 = (\frac{1}{\rho} \beta_{X_2} \odot X_4 + X_4 \odot X_2 - \frac{1}{\rho} \beta_{X_3} + X_5 ) / (X_4 \odot X_4 + 1)\]
\[X_4 = (\frac{1}{\rho} \beta_{X_2} \odot X_3 + X_3 \odot X_2 - \frac{1}{\rho} \beta_{X_4} + X_6 ) / (X_3 \odot X_3 + 1)\]
\[X_5 = \text{algorithm used for Markov-State term above, using a }\]
\[ \text{modified cost matrix $\tilde{C}_k$, see below} \quad\quad\quad\quad\quad\]
\begin{flalign*}
X_6(k,\ell,n) & := \begin{cases}
w(k,\ell,n), & \minact < w(k,\ell,n) \\
\minact & \frac{\minact}{2} \leq w(k,\ell,n) \leq \minact\\
0 & \text{otherwise}
\end{cases} & 
\end{flalign*}
Here, $\dict^\top$ denotes the adjoint of $\dict$, the division is element-wise, the shorthands $w(k,\ell,n) := \frac{1}{L} \sum^L_{\tilde{\ell}=1} u(k,\tilde{\ell},n+\ell-\tilde{\ell})$, $u := X_4 + \frac{1}{\rho} \beta_{X_4}$ and $v := X_3 + \frac{1}{\rho} \beta_{X_3}$, as well as the cost matrix $\tilde{C}_k$
\begin{equation}\tilde{C}_k(\ell,n) = \sum_{\mathclap{\substack{\tilde{\ell} \in \{1,\ldots,L\} \setminus \\ \{\ell,\ldots,\ell+\lceil w/s\rceil\}}}} v(k,\tilde{\ell},n)^2
+
\sum_{\mathclap{\tilde{\ell} \in \{\ell,\ldots,\ell+\lceil w/s\rceil\}}} (v(k,\tilde{\ell},n) - 1)^2
\label{eq:costMatrixBMS}
\end{equation}
After the minimization, we maximize over the dual variables (as we have done in ADMM):
\[
\beta_{X_1} = \beta_{X_1} + \rho (X_1- \apspecvar{X_2}) \qquad
\beta_{X_2} = \beta_{X_2} + \rho (X_2 - X_3 \odot X_4)
\]
\[
\beta_{X_3} = \beta_{X_3} + \rho (X_3-X_5) \qquad\;\;\,
\beta_{X_4} = \beta_{X_4} + \rho (X_4-X_6) \qquad\qquad\qquad
\]
Iterating these updates, $X_5$ and $X_6$ contain estimates for $\actB$ and $\actG$.

\section{Experiments}
\label{sec:Experiments}

\begin{table}[t]
\caption{Description of dataset UM-NI.}
\label{tab:dataset_UM-NI}
	\centering
\begin{tabular}{llllccc}
\textbf{ID} & \textbf{Composer} & \textbf{Piece} & \textbf{Performer}\\
\hline
U01 & Bach & BWV.~851 & Colafelice \\
U02 & Beethoven & Op.~10 No.~3 & Wang\\
U03 & Chopin & Op.~25 No.~11 & Kim  \\
U04 & Haydn & HobXVI 52 & Mizumoto\\
U05 & Liszt & Polonaise E-Maj & Denisova\\
U06 & Mendelssohn & Op.~54 & Sham  \\
U07 & Mozart & K284-01 & Ozaki \\
U08 & Ravel & Alb. D. Grac. & Teo \\
U09 & Schubert & Op.~142 No.~3 & Chon  \\
U10 & Stravinsky & Op.~7 No.~4 & Lin \\
\end{tabular}
\end{table}

To illustrate the performance of our proposed method, we conducted a series of experiments. 
Since our method relies on the existence of individual recordings for each piano key, the set of available datasets is more limited than in most transcription scenarios. Overall, we use three datasets, each having specific properties. 

\subsection{University of Minnesota / Native Instruments (UM-NI) Dataset}
\label{subsec:UMNI}

The first dataset was also used in \cite{EwertPS15_DPNMD_ICASSP} and
comprises ten MIDI files downloaded from the University of Minnesota piano-e-competition website\footnote{\url{http://www.piano-e-competition.com}}.
These were recorded using a Yamaha Disklavier during an international piano playing competition
and therefore closely capture the actual, real-world performance of skilled pianists. 
The pieces were selected to cover a broad range of composers and performers but were otherwise selected randomly, see Table~\ref{tab:dataset_UM-NI} for an overview.
To create high quality audio versions from these MIDI files, we employed Native Instruments' Vienna Concert Grand VST plugin, which comprises samples of a Boesendorfer 290 concert grand with an uncompressed size of almost 14 GB. Additionally, we used the plugin to create recordings of single notes for that piano, each 6 seconds long and played in mezzo forte (MIDI velocity 75).
We refer to this dataset in the following as \emph{UM-NI} and use it to investigate the influence of parameters on our proposed method and to compare our results with the system presented in \cite{EwertPS15_DPNMD_ICASSP}.

To evaluate a method, we employ precision (P), recall (R), and F-measure (F) values as used in the MIREX evaluation campaigns.
A detected note is considered correct if there is a note in a corresponding ground truth MIDI
file having the same MIDI pitch, with on onset position up to 50ms apart from the detected note. Every ground truth note can only validate
up to one detected note. By counting the number of correctly detected notes (TP), incorrectly detected extra notes (FP) and incorrectly missed notes (FN), we can define the precision P := TP / (TP + FP), recall R := TP / (TP + FN) and f-measure F := 2 $\cdot$ P $\cdot$ R / (P + R).

\begin{figure}
		\hspace{-0.2cm}
    \begin{tikzpicture}[scale=1.0, every node/.style={transform shape}]
			\node[anchor=south west] (label) at (0.0,0.15){\includegraphics[width=8.7cm]{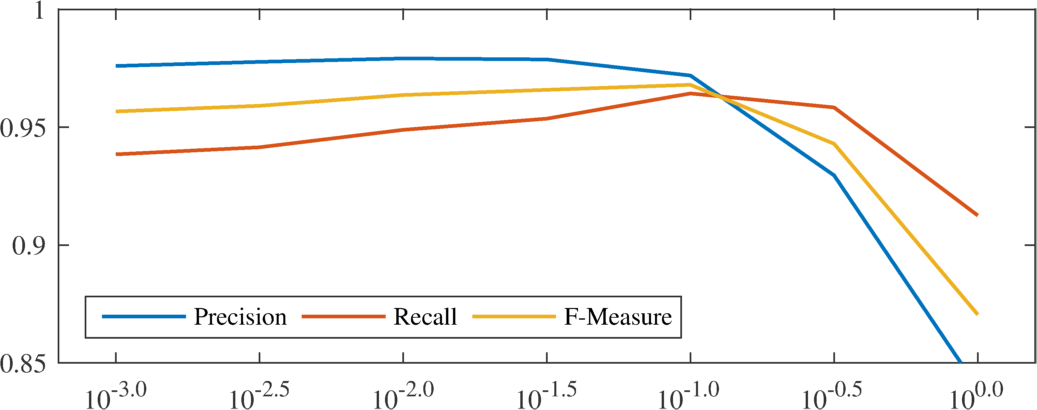}};
			\node at (4.6,0) {\footnotesize \textcolor{labelgray}{Parameter $\lambda_1$}};
    \end{tikzpicture}
		\vspace{-0.6cm}
    \caption{Averaged evaluation results using the UM-NI dataset for various values of the sparsity parameter $\lambda_1$.}
		\label{fig:experiment_alpha1}
\end{figure}

\begin{figure}
		\hspace{-0.2cm}
    \begin{tikzpicture}[scale=1.0, every node/.style={transform shape}]
			\node[anchor=south west] (label) at (0.0,0.15){\includegraphics[width=8.7cm]{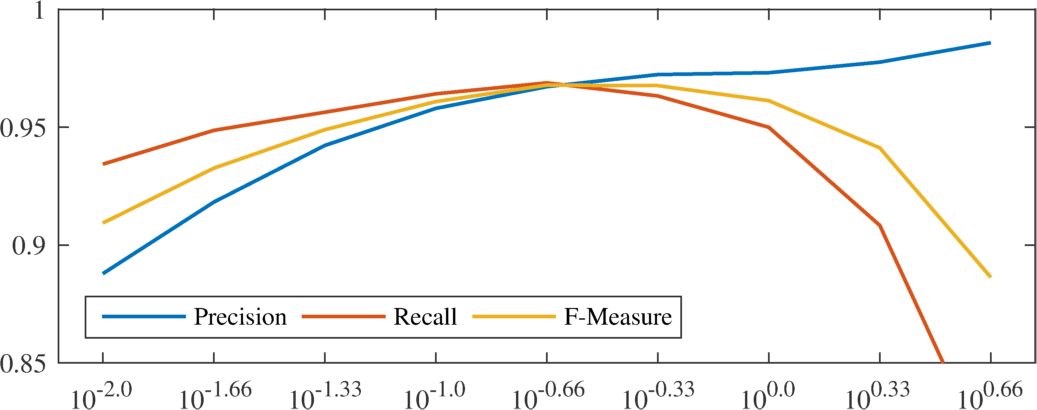}};
			\node at (4.6,0) {\footnotesize \textcolor{labelgray}{Parameter $\lambda_2$}};
    \end{tikzpicture}
		\vspace{-0.6cm}
        \caption{Averaged evaluation results using the UM-NI dataset for various values of the total diagonal variation parameter $\lambda_2$.}
		\label{fig:experiment_alpha2}
\end{figure}

Overall, our proposed system has two main parameter for sparsity $\lambda_1$ and total diagonal variation $\lambda_2$. 
Informal tests using the UM-NI showed a good performance setting $\lambda_1 = 0.1$ and $\lambda_2 = 0.5$. With our first two experiments we investigate the effect these two parameters have in more detail. We first set up the dictionary tensor $\dict$ using the single note recordings.
Then, we use our system (without the binary Markov-state and strict coupling regularizer for now) to obtain an activation value for a middle C played pianississimo (very softly) -- this activation value minus 10\% is used to set the minimum activation value $\minact$. After estimating $\act$ using our proposed method ($300$ iterations minimizing $\tobjf_d$, which is used as initialization for $300$ iterations with $\tobjf_f$), we detect an onset for key $k$ in frame~$n$ if $\act(k,1,n) \ge \minact$ and $\act(k,1,n)\ge\act(k,1,\tilde{n})$
for all $\tilde{n}\in\{n-L_M,\ldots,n+L_M\}$.
\figurename~\ref{fig:experiment_alpha1} shows our evaluation results for various values of $\lambda_1$ fixing $\lambda_2=0.5$, and 
\figurename~\ref{fig:experiment_alpha2} for various values of $\lambda_2$ fixing $\lambda_1 = 0.1$.
In \figurename~\ref{fig:experiment_alpha1} we can observe that lower values for $\lambda_1$ actually yield higher precision values, which is counter-intuitive at first as a higher weight for the sparsity term typically yields fewer and more meaningful activations. The underlying cause here is the use of a dictionary containing unnormalized templates, as already discussed in Section~\ref{sec:Model}.
Further, we can see that the f-measure is reasonably stable for $\lambda_1 \le 10^{-1}$. The break-even point is around $10^{-1}$, which also coincides with a small local maximum for the f-measure. Therefore, we keep $\lambda_1 = 0.1$ in the following. 
For $\lambda_2$, the f-measure is quite stable and above $0.94$ with $\lambda_2 \in [10^{-1.33},10^{0.33}]$. The break-even point and the maximum of the f-measure are between $10^{-0.66}$ and $10^{-0.33}$, which made us adjust the $\lambda_2$ value slightly to $0.4$ in the following. Evaluation results for each piece in UM-NI using these settings are shown in Table~\ref{tab:UMNIResults}, which enables a comparison with the method presented in \cite{EwertPS15_DPNMD_ICASSP}. As discussed in Section~\ref{sec:Related}, this system is a hybrid between NMD and FS-HMM and employs the same spectro-temporal patterns we use as internal dictionary. As we can see, the proposed system clearly outperforms \cite{EwertPS15_DPNMD_ICASSP} on every recording.
An informal manual investigation showed that the reason for this considerable difference was in many cases that the method proposed in \cite{EwertPS15_DPNMD_ICASSP} generated note object candidates for a given pitch by actually modelling energy for another pitch -- an effect resulting from the decoupled parameter estimation (as discussed at the end of Section~\ref{sec:Related}). Our proposed method does not decouple parameters for different pitches during the important first step using convex terms and thus is less prone to run into less meaningful local minima overall.

\begin{table}[t]
	\caption{Evaluation results: Precision, Recall and F-Measure for detected note onsets for UM-NI dataset.}
	\label{tab:UMNIResults}
	\centering
\setlength{\tabcolsep}{1.0ex}
\begin{tabular}{ll|ccccccccccc}
 & & \textbf{U01} & \textbf{U02} & \textbf{U03} & \textbf{U04} & \textbf{U05} & \textbf{U06} & \textbf{U07} & \textbf{U08} & \textbf{U09} & \textbf{U10} & \textbf{Av}\\
\hline
\parbox[t]{2mm}{\multirow{3}{*}{\rotatebox[origin=c]{90}{\cite{EwertPS15_DPNMD_ICASSP}}}}
&P& 91 & 86 & 76 & 86 & 90 & 94 & 83 & 86 & 92 & 84& \textbf{87}\\
&R& 93 & 89 & 83 & 87 & 92 & 84 & 89 & 87 & 91 & 96& \textbf{89}\\
&F& 92 & 87 & 79 & 86 & 91 & 89 & 86 & 86 & 91 & 89& \textbf{88}\\
\hline
\parbox[t]{2mm}{\multirow{3}{*}{\rotatebox[origin=c]{90}{Prop.}}}
&P& 100& 97& 94& 98& 96& 98& 98& 97& 98& 95& \textbf{97}\\
&R& 100& 97& 94& 97& 93& 94& 96& 98& 100& 99& \textbf{97}\\
&F& 100& 97& 94& 97& 95& 96& 97& 97& 99& 97& \textbf{97}\\
\end{tabular}
\end{table}

\subsection{MIDI Aligned Piano Sounds (MAPS) Dataset}

While the UM-NI dataset contains MIDI files of real performances, the corresponding audio recordings are synthesized, which can have a great influence on evaluation results. Therefore, we conducted additional experiments using the MAPS dataset \cite{EmiyaBD10_MultipitchEstimation_TASLP}, which contains audio recordings of MIDI files played on a Yamaha Disklavier. In contrast to UM-NI, MAPS contains only score-like MIDI files and is divided into two subsets, which correspond to different microphone placements. For the ENSTDkCl subset a close miking configuration was used, while the ENSTDkAm recordings were made using an ambient configuration. Due to room acoustics, the latter contains a considerable amount of reverberation and thus is generally considered as the more difficult to transcribe. 

\begin{figure}
		\hspace{-2mm}
		\centering
    \begin{tikzpicture}[scale=1.0, every node/.style={transform shape}]
			\begin{scope}[shift={(0.05,2.7)}]
			\node[anchor=south west] (label) at (0,0){\includegraphics[width=8.45cm]{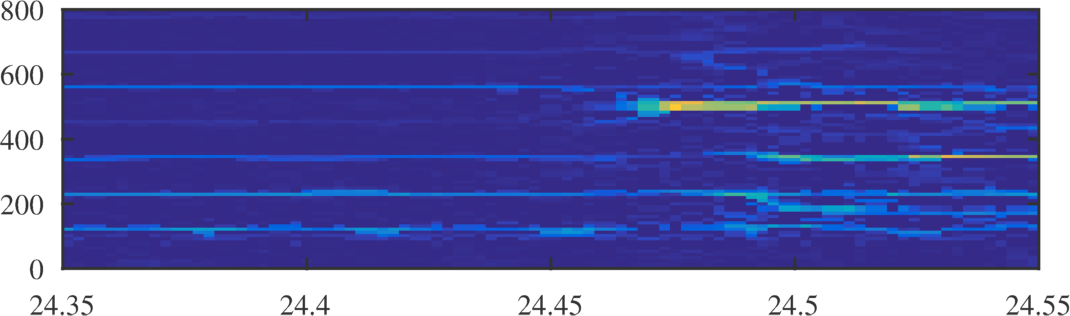}};
			\end{scope}
			\begin{scope}[shift={(0,0)}]
      \node[anchor=south west] (label) at (0,0){\includegraphics[width=8.5cm]{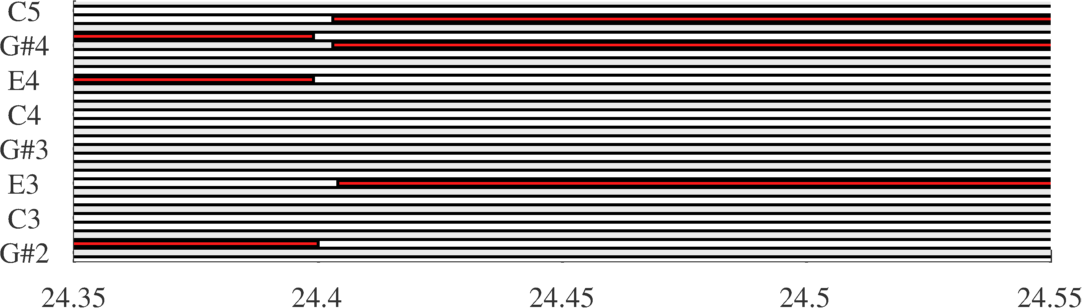}};
			\node at (4.6,-0.1) {\scriptsize \textcolor{labelgray}{Time [sec]}};
			\draw[annotred,line width=0.4mm,,rounded corners=2pt] (5.0,4.3) rectangle (5.6,4.65);
			\draw[annotred,line width=0.4mm,,rounded corners=2pt] (5.9,3.95) rectangle (6.5,4.2);
			\draw[annotred,line width=0.4mm,,rounded corners=2pt] (6.0,3.55) rectangle (6.6,3.8);
			\end{scope}
    \end{tikzpicture}
		\caption{Spectrogram for a section of the recording \emph{MAPS\_MUS-mendel\_op62\_5\_ENSTDkAm} and corresponding annotation MIDI file as contained in the MAPS dataset. Markers indicate onset positions in the recording.} 
		\label{fig:MAPSProb}
\end{figure}

For both datasets, we decided to provide results using two different temporal tolerances: one using the usual $\pm$50ms and one using $\pm$100ms. We include the greater tolerance as we found that the actual temporal accuracy of the MIDI-based annotations included in MAPS sometimes exceed the documented accuracy of 20ms by an order of magnitude.
This is illustrated in Fig.~\ref{fig:MAPSProb}, where we see three notes of an E-major chord that are supposed to start concurrently at $24.404$sec in the MIDI file. In the recording, however, the corresponding onsets are approximately at $24.47$sec (B), $24.49$sec (G$^\sharp$) and $24.495$sec (E). Overall, in this example, the difference is $65$-$90$ms. Manually inspecting the dataset using our own MIDI parser and verifying the findings using the Audacity audio editor, we observed that these differences are often not simple offsets. Rather, some seem to be caused by slow drifts which could be related to differences in clock speed between the MIDI and the audio equipment. More importantly though we also observed a considerable temporal jitter which, for instance, leads to concurrent notes being played asynchronously in the recording (as in Fig.~\ref{fig:MAPSProb}). This could be related to a limitation of the Disklavier series and player pianos in general. In particular, when a Disklavier is controlled via direct MIDI input, the instrument plays each note as quickly as possible to minimize delays. Depending on the piano key and the velocity, however, each hammer needs a different amount of time to hit the strings, which consequently can lead to different onset times in a recording. Since the MAPS documentation suggests that the recordings were made using direct MIDI input, this could explain the apparent temporal jitter we observed. Therefore, we believe that the greater temporal tolerance might provide a more realistic impression of the transcription performance. 

\begin{table}[t]
	\caption{Precision, Recall and F-Measure in percent for various methods. The results are given separately for the two subsets recorded using a Yamaha Disklavier as provided by the MAPS dataset.}
	\label{tab:dataset_MAPS}
	\centering
	\scalebox{1.1}{
\begin{tabular}{clccc}
\textbf{Subset} & \textbf{Method} & \textbf{P} & \textbf{R} & \textbf{F}\\
\parbox[t]{2mm}{\multirow{7}{*}{\rotatebox[origin=c]{90}{\small Close}}}
& Benetos/Ewert/Weyde \cite{BenetosEW14_PitchUnpitchedTranscription_ICASSP} & --- & --- &  63 \\
& FitzGerald/Cranitch/Coyle \cite{FitzGeraldCC08_ExtendedNMFVariants_CIN} &  60 & 59 & 58 \\
& O'Hanlon/Nagano/Plumbley \cite{OhanlonNP12_StructuredSparsityAMT_ICASSP} & 79 & 77 & 78 \\
& Tolonen/Karjalainen \cite{TolonenK00_EfficientMultipitch_TSAP} & 62 & 22 & 32 \\
& Vincent/Bertin/Badeau \cite{VincentBB10_AdaptoveHarmonicNMF_TASLP}  & 72 & 66 & 67 \\
& \textbf{Proposed ($\pm$50ms)} & \textbf{91} & \textbf{88} & \textbf{89} \\  %
& \textbf{Proposed ($\pm$100ms)} & \textbf{96} & \textbf{93} & \textbf{95} \\ %
\hline
\parbox[t]{2mm}{\multirow{7}{*}{\rotatebox[origin=c]{90}{\small Ambient}}}
& Bertin/Badeau/Vincent \cite{BertinBV10_EnforcingHarmonicityInBayesNMF_TASLP} & 47 & 45 & 45 \\
& Carabias-Orti et al. \cite{CarabiasVVRC11_MultiInstrExcitation_JSTSP}  & --- & --- & 52 \\
& Klapuri \cite{Klapuri06_multiF0_ISMIR}  & --- & --- & 55 \\
& Marolt \cite{Marolt04_AMT-NN-Piano_TMM}  & 64 & 54 & 58 \\
& Virtanen \cite{Virtanen07_MonauralSoundSourceSeparation_TASLP}  & 34 & 35 & 34 \\
& \textbf{Proposed ($\pm$50ms)} & \textbf{86} & \textbf{85} & \textbf{85} \\  %
& \textbf{Proposed ($\pm$100ms)} & \textbf{93} & \textbf{92} & \textbf{93} \\
\end{tabular}
}
\end{table}

The results for our proposed method as well as various previously published methods (\cite{OhanlonNP12_StructuredSparsityAMT_ICASSP,BenetosEW14_PitchUnpitchedTranscription_ICASSP,VincentBB10_AdaptoveHarmonicNMF_TASLP,BertinBV10_EnforcingHarmonicityInBayesNMF_TASLP,CarabiasVVRC11_MultiInstrExcitation_JSTSP}) are given in Table~\ref{tab:dataset_MAPS}, with our method using the parameters found using the UM-NI dataset as discussed above.
We remark that many of these methods do not use single note recordings as training material as we do, and thus the numbers typically cannot fairly be  compared with our method. Some general observations, however, might be possible. In particular, the considerable jump in f-measure from previous methods might indicate that our proposed method modeling both spectral and temporal signal properties could be capable of exploiting the provided prior knowledge in the form of single note recordings to a high degree. Even under reverberated conditions (ambient dataset in Table~\ref{tab:dataset_MAPS}), the drop in performance for our method (from $95$\% f-measure to $93$\%) is less pronounced than for the remaining methods (average f-measure for the close subset is $60$\% and $49$\% for the ambient subset). Overall, combining the results from the UM-NI and the MAPS subsets could potentially indicate that our proposed method can be used to obtain transcription results of high accuracy even under real world conditions. 

\subsection{Influence of Individual Regularizers and Error Analysis}

To indicate the influence of each regularizer in our method, we conducted a series of experiments using the close-miking subset of the MAPS collection -- the ambient subset showed a similar behavior. To this end, we start with our objective function $\tobjf_b$, which contains only the Kullback-Leibler and the non-negativity terms. By adding the remaining regularizers successively, we illustrate the relative change in performance.
A challenge, however, is that our proposed method contains components inside its parameter estimation procedure that most previous methods implement as an additional post-processing step. For example, many methods employ a Hidden Markov Model to decode NMF activations into a set of note objects \cite{KlapuriD06_SPforMusic_BOOK} -- a task undertaken by our Markov-State regularizer inside the parameter estimation. 
Therefore, we provide two different results: one using the same decision logic as used before (essentially binarization, compare Section~\ref{subsec:UMNI}) and one using an HMM that decodes note objects from the final activations similar to existing methods \cite{KlapuriD06_SPforMusic_BOOK} (but following the graphical model shown in Fig.~\ref{fig:graphModelSoundProgression} to account for the fact that we use a structured dictionary tensor and not a flat matrix). The results are shown in Table.~\ref{tab:results_individualRegs}.

\begin{table}[t]
	\caption{F-Measure in percent for detected note onsets for for several variants of our proposed method eliminating individual regularizers using the close-miking subset of the MAPS dataset. The \emph{Direct} and \emph{HMM} columns refer to different methods for producing the final note objects as described in the text.}
	\label{tab:results_individualRegs}
	\centering
	\scalebox{1.1}{
\begin{tabular}{lcc}
\textbf{Regularizer} & \textbf{Direct} & \textbf{HMM} \\
 Kullback-Leibler \& Non-Negativity  & 31 & 56  \\  
 + Sparsity  & 74 & 81 \\ 
 + Total Diagonal Variation  & 90 & 91  \\ 
 + Markov-State Reg. & 94 & 94 \\ 
 + Thresholding Reg.  & 95 & 95  \\ 
 + Bin.~Markov-State \& Strict Coupling Reg. & 92 & 93  \\ 
\end{tabular}
}
\end{table}

As we can see, using only the Kullback-Leibler divergence and the non-negativity term results are significantly worse ($31$/$56$\%) -- worse than many of the methods shown in Table~\ref{tab:dataset_MAPS} despite our use of instrument-specific single note recordings. Here, without any additional regularizers, the resulting activations are often unstructured and contain many spurious entries. In particular, onset sounds for notes with velocities vastly different from the ones used for the single notes are synthesized using what seems like random combinations of templates from unrelated piano keys. Including the sparsity term improves the performance considerably ($74$/$81$\%). As already discussed in Section~\ref{subsec:sparsity}, the sparsity term alone already eliminates many spurious activations but also introduces additional ones due to our use of unnormalized templates. Another considerable jump in performance is the result of using our TDV term ($90$/$91$\%) -- this term is particularly useful due to our approach of using unnormalized dictionaries which already capture the amplitude progression of a piano sound over time, including any non-stationary changes in spectral properties. This leads to very little fluctuation of activity over time and thus we can set the relative importance of the TDV term very high. The TDV term also leads to similar semantics as Markov terms, which is possible due to the Bakis-type structure in our graphical model -- however, in a completely convex formulation eliminating any numerical problems related to usual Markov-type constraints. This is evident in the fact that the 'HMM' value is not much higher than the 'Direct' f-measure.

With respect to the additional non-convex terms, including the Markov constraints within instead of outside the parameter estimation introduces additional harder semantics, which eliminate some of the remaining uncertainties ($94$/$94$\%). Further, including the thresholding term inside the parameter estimation allows explaining some energy in the signal which would remain unexplained otherwise, which leads to another small improvement ($95$/$95$\%). Finally, including the binary Markov-state and strict coupling regularizers reduces again the onset f-measure, as already discussed in Section~\ref{subsec:BinMarkovStrictCoupl}. Here the reason is that the strict coupling of activations (similar to NMD in concept) is too restrictive and can lead to additional, incorrect note detections -- the usefulness of these terms thus remains limited to the estimation of note lengths.

Overall, one might argue that the non-convex terms do not considerably increase the performance anymore compared to the results obtained from the convex terms alone, with the f-measure only increasing from $90$\% to $95$\%. However, this small change means that we actually halved the number of incorrect notes in our result using the non-convex terms and only expect five instead of ten wrong notes in a hundred notes. 

In another experiment, we investigate the capability of our method to additionally compute the duration of each note. As evaluation measure, we use the procedure used in MIREX, i.e. a detected note is considered as correct only if the onset is correct as described above and the detected note duration is within $20$\% of the corresponding note-length in the ground truth MIDI file. To obtain the correct note length from a given ground truth MIDI file, we parsed the events related to the sustain pedal in each MIDI file and adjusted the note lengths accordingly. 
We evaluate our proposed method in two configurations: first, using the activity $\act$ obtained by minimizing $\tobjf_d$ followed by $\tobjf_f$, and second, using $\actB$ and $\actG$ resulting from a minimization of $\tobjf_g$, with the solution initialized based on the $\act$ obtained from the first step (compare Section~\ref{sec:NoteLengthModel}). Using the resulting $\act$ and $\actB$, we determine a duration for each onset, which we detect as before. More precisely, for a detected onset for key $k$ in frame $n$, we find $\ell$ with $\act(k,\ell,n+\ell-1)\ge \minact$ and $\act(k,\ell+1,n+\ell)<\minact$ and set the duration for the note to the time in seconds corresponding to $\ell$ frames. For the second method, we do the same but find $\ell$ with
$\actB(k,\ell,n+\ell-1)=1$ and $\actB(k,\ell+1,n+\ell)=0$.
Using a tolerance of $\pm$100ms for the onset, we compute precision, recall and f-measure using the duration-based error measure.
For the first method, using the close miking and ambient subsets in MAPS, we obtain an f-measure of $46$\% and $38$\%, respectively.
Using the second method, we yield an f-measure of $59$\% and $55$\%, respectively. This illustrates, that while using $\tobjf_g$ for onset detection can decrease the performance due to unrealistic assumptions expressed by the regularisers, it can improve the note length estimation by a considerable amount.

As a final step, we manually tried to identify some general sources of error to find out why and when our method failed.
First, our approach of using the same threshold for all piano keys, and choosing this threshold based on a single note, was sometimes not precise enough. Therefore, the performance of our method could be improved if the minimum activity $\minact$ would be available for more piano keys, for example by providing additional recordings of other softly played notes. 
As a second cause, when the sustain pedal is used, some strings are cross-excited when strings in a neighborhood are hit -- depending on the velocity and instrument specific factors. It is a matter of definition if such detections should indeed be counted as transcription errors. Third, depending on the playing style and the velocity in particular, the temporal decay rates for partials and the spectral envelope change sometimes drastically -- with considerable differences between different instruments and also for individual keys on the same piano. Since we use only one time-frequency pattern for each key, this can lead to pattern mismatches, leaving residual energy which in some cases is modeled using wrong patterns. These pattern mismatches had a stronger effect under reverberant conditions (which explain the lower results for the ambient subset). 

We also investigated sources of error for the note length detection. First, the duration for softly played notes is more often incorrect than for forte note, which is caused by a difference in the energy decay between our mezzo-forte pattern and the actual note. Second, the use of the sparsity term during the initialization can lead to activation diagonals that are too short (as later templates in a pattern contain less energy and are not activated anymore). While the sparsity term is not used in $\tobjf_g$, it is used to obtain an initialization and we observed that the update rules presented in Section~\ref{sec:NoteLengthModel} cannot always recover from this error. Third, after the offset, the string is still vibrating for a short while (release state in ADSR model) introducing a principal temporal uncertainty where the actual note end is.

\section{Conclusions}
\label{sec:Conclusions}

We presented a method for transcribing pitched-percussive instruments such as the piano in controlled recording conditions.
Compared to NMF, where time and frequency are strictly separated, our method employs spectro-temporal patterns to model the temporal evolution for each individual piano key.
In contrast to non-negative matrix deconvolution, these patterns can be of variable length if activated.
From a numerical point of view, our approach employs a combination of convex and non-convex regularizers, which penalize unwanted behavior in an otherwise loosely defined signal model.
This is in contrast to the FS-HMM and similar approaches, where the temporal evolution of spectra is directly enforced by a graphical model. 
The result is a highly accurate parameter estimation which does not rely on unstable parameter decoupling techniques and thus is less prone to poor local minima. 
The optimization framework ADMM that we used is highly extensible and should be useful in many other audio-processing scenarios.
Overall, the combination of prior knowledge available in controlled recording conditions with our proposed signal model seems to provide a considerable boost in transcription performance.
For the future it could be interesting to post-process the output of our method using discriminative methods -- this could combine the best of two worlds, as our method provides a straightforward integration of prior knowledge in the form of single notes and thus a straightforward adaptability to new acoustic environments, while discriminative methods such as RNN-based language models \cite{SigtiaBCWGD14_RNNLanguageModelTranscription_ISMIR} capture higher level semantics and musical expectation.

\section*{Acknowledgement}

This work was supported by the
Engineering and Physical Sciences Research Council (EPSRC) programme EP/L019981/1.

\bibliographystyle{IEEEtran}
\bibliography{referencesMusic}

\begin{IEEEbiography}[{\includegraphics[width=1in,height=1.25in,clip,keepaspectratio]{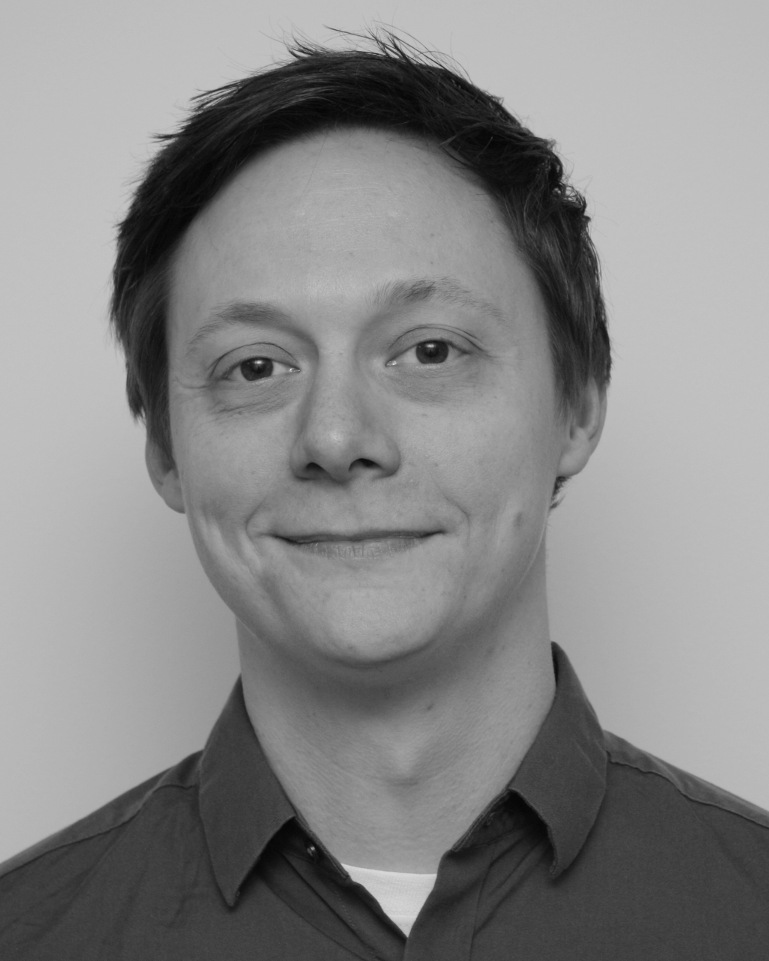}}]{Sebastian Ewert}
received the M.Sc./Diplom and Ph.D. degrees (summa cum laude) in computer science from the University of Bonn (svd. Max-Planck-Institute for Informatics), Germany, in 2007 and 2012, respectively. 
After a postdoc at the Centre for Digital Music, Queen Mary University of London (United Kingdom), he became lecturer for signal processing in the centre in 2015. Currently, he is additionally holding a research position in the EPSRC programme Fusing Audio and Semantic Technologies (FAST) and is one of the leaders of the Machine Listening Lab. 
\end{IEEEbiography}

\begin{IEEEbiography}[{\includegraphics[width=1in,height=1.25in,clip,keepaspectratio]{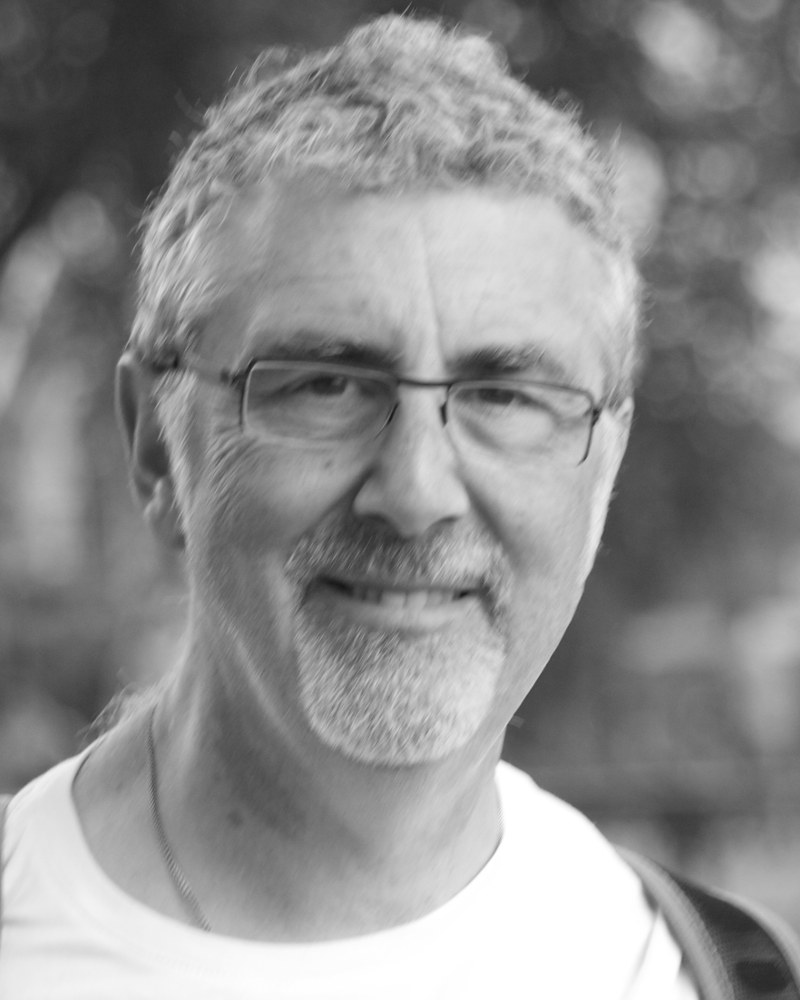}}]{Mark Sandler}
(PhD FAES FIET FIEEE CEng) is Founding Director of the Centre for Digital Music, a world-leading research group in audio and music technology with over 80 members. The Centre is in the School of Electronic Engineering \& Computer Science, where he holds the chair in Signal Processing. He is also Director of the Centre for Doctoral Training in Media and Arts Technology, a UK government funded special PhD training programme. He is a recipient of the Royal Society Wolfson Research Merit Award (2015-19). He has published over 400 papers in conferences and journals.  
\end{IEEEbiography}

\vfill

\end{document}